\documentclass[12pt,a4paper]{article}
\usepackage{a4wide}
\usepackage{amsmath}
\usepackage{amssymb}
\usepackage{txfonts}
\begin{document} 
{\renewcommand{\thefootnote}{\fnsymbol{footnote}}
\mbox{}\hfill FSUJ-TPI-05/03\\
\mbox{}\hfill CGPG--03/4--5\\
\medskip
\begin{center} {\LARGE Classical Solutions for
 Poisson Sigma Models \\ on a Riemann surface}\\
\vspace{1.5em} Martin Bojowald$^a$\footnote{e-mail
address: {\tt bojowald@gravity.psu.edu}} and Thomas
Strobl$^b$\footnote{e-mail address: {\tt
Thomas.Strobl@tpi.uni-jena.de}}
\\\vspace{0.5em} $^a$Center for Gravitational Physics and Geometry,
Department of Physics,\\ The Pennsylvania State University, University
Park, PA 16802, USA\\\vspace{0.5em} $^b$Institut f\"ur
Theoretische Physik, Universit\"at Jena, D--07743 Jena, Germany\\
\vspace{1.5em}
\end{center} }

\def\2{{\textstyle\frac{1}{2}}}
\def\ba{\begin{eqnarray}}
\def\ea{\end{eqnarray}}
\def\be{\begin{equation}}
\def\ee{\end{equation}}
\def\re{(\ref }
\newcommand{\Kern}{\mathop{\mathrm{ker}}}
\newcommand{\rank}{\mathop{\mathrm{rank}}}
\newcommand{\Ad}{\mathop{\mathrm{Ad}}}
\newcommand{\ad}{\mathop{\mathrm{ad}}\nolimits}
\newcommand{\tr}{\mathop{\mathrm{tr}}}
\newcommand{\Mod}{{\CM}_{\mathrm{cl}}}
\newcommand{\fra}{\mathfrak{a}}
\newcommand{\frg}{\mathfrak{g}}

\let\a=\alpha \let\b=\beta \let\g=\gamma \let\d=\delta
\let\e=\varepsilon \let\ep=\epsilon \let\z=\zeta \let\h=\eta
\let\th=\theta
\let\dh=\vartheta \let\k=\kappa \let\l=\lambda \let\m=\mu
\let\n=\nu \let\x=\xi \let\p=\pi \let\r=\rho \let\s=\sigma
\let\t=\tau \let\o=\omega \let\c=\chi \let\ps=\psi
\let\ph=\varphi \let\Ph=\phi \let\PH=\Phi \let\Ps=\Psi
\let\O=\Omega \let\S=\Sigma \let\P=\Pi
\let\Th=\Theta \let\L=\Lambda \let \G=\Gamma \let\D=\Delta
\def\wtO{\widetilde{\Omega}}

\def\({\left(} \def\){\right)} \def\<{\langle} \def\>{\rangle}
\def\lb{\left\{} \def\rb{\right\}}
\let\lra=\leftrightarrow \let\LRA=\Leftrightarrow
\def\ul{\underline}
\def\wt{\widetilde}
\let\Ra=\Rightarrow \let\ra=\rightarrow
\let\la=\leftarrow \let\La=\Leftarrow

\def\CG{{\cal G}}\def\CN{{\cal N}}\def\CC{{\cal C}}
\def\CL{{\cal L}} \def\CX{{\cal X}} \def\CA{{\cal A}} \def\CE{{\cal
E}}
\def\CF{{\cal F}} \def\CD{{\cal D}} \def\rd{\rm d}
\def\rD{\rm D} \def\CH{{\cal H}} \def\CT{{\cal T}} \def\CM{{\cal M}}
\def\CI{{\cal I}}
\def\CP{{\cal P}} \def\CS{{\cal S}} \def\C{{\cal C}}
\def\CR{{\cal R}}
\def\CO{{\cal O}}
\def\CU{{\cal U}}

\newcommand*{\dR}{{\mathbb R}}
\newcommand*{\dN}{{\mathbb N}}
\newcommand*{\dZ}{{\mathbb Z}}
\newcommand*{\dC}{{\mathbb C}}
\newcommand{\md}{\mathrm{d}}
\newcommand{\mdp}{\mathop{{\mathrm{d}}_{\parallel}}\nolimits}
\newcommand{\mdo}{\mathop{{\mathrm{d}}_{\perp}}\nolimits}
\newcommand{\Diff}{\mbox{\rm Diff}}
\newcommand{\diff}{\mbox{\rm diff}}

\makeatletter
\def\bard{\protect\@bard}
\def\@bard{%
\relax
\bgroup
\def\@tempa{\hbox{\raise.73\ht0
\hbox to0pt{\kern.4\wd0\vrule width.7\wd0
height.1pt depth.1pt\hss}\box0}}%
\mathchoice{\setbox0\hbox{$\displaystyle\mathrm{d}$}\@tempa}%
{\setbox0\hbox{$\textstyle \mathrm{d}$}\@tempa}%
{\setbox0\hbox{$\scriptstyle \mathrm{d}$}\@tempa}%
{\setbox0\hbox{$\scriptscriptstyle \mathrm{d}$}\@tempa}%
\egroup
}
\makeatother

\def\Rboost{\dR}

\def\oG{\hbox{${\cal G}_+^\uparrow$}}

\newtheorem{theo}{Theorem}
\newtheorem{lemma}{Lemma}
\newtheorem{cor}{Corollary}
\newtheorem{defi}{Definition}
\newtheorem{prop}{Proposition}

\newcommand{\proofend}{\raisebox{1.3mm}{%
\fbox{\begin{minipage}[b][0cm][b]{0cm}\end{minipage}}}}
\newenvironment{proof}[1][\hspace{-1mm}]{{\noindent\it Proof #1:}
}{\mbox{}\hfill \proofend\\\mbox{}}

\begin{abstract}
We determine the moduli space of classical solutions to the field
equations of Poisson Sigma Models on arbitrary Riemann surfaces for
Poisson structures with vanishing Poisson form class. This condition
ensures the existence of a presymplectic form on the target Poisson
manifold which agrees with the induced symplectic forms of the Poisson
tensor upon pullback to the leaves. The dimension of the classical
moduli space as a function of the genus of the worldsheet $\S$ and the
corank $k$ of the Poisson tensor is determined as $k \,
(\rank(H^1(\S))+1)$.  Representatives of the classical solutions are
provided using the above mentioned presymplectic 2-forms, and possible
generalizations to cases where such a form does not exist are
discussed. The results are compared to the known moduli space of
classical solutions for two-dimensional BF and Yang--Mills theories.
\end{abstract}

\setcounter{footnote}{0}

\section{Introduction} Poisson Sigma Models (PSMs) \cite{PSM1,Ikeda}
are topological or almost topological two-dimensional field theories
associated to a Poisson manifold. Given any Poisson bracket on a
manifold $M$, characterized
by a Poisson bivector $\CP=\frac{1}{2} \CP^{ij}(X) \partial_i \wedge
\partial_j$ where $X^i$, $i=1,\ldots,n$, are local coordinates on $M$,
the topological part of the action has the form
\be S = \int_\S A_i \wedge \md X^i + \2 \CP^{ij} A_i \wedge
A_j \, .
\label{action} \ee It is a functional of the fields $X(x)$, which
parametrize a map $\CX$ from the two-dimensional worldsheet $\S$ into
the target $M$, as well as 1-forms on the worldsheet $\S$ taking
values in the pullback of $T^\ast M$ by the map $\CX$, $A_i =
A_{i\mu}\md x^\mu$, $\mu = 1,2$. More compactly, $S$ may be regarded
as a functional on the vector bundle morphisms from $T\Sigma$ to
$T^*M$ \cite{CFHam}.
The action remains topological if, e.g., one adds to it the pullback
of a 2-form $B$ which has an exterior derivative $H=\md B$ that
vanishes upon contraction with any Hamiltonian vector field of $\CP$
(cf \cite{Ctirad} for further details as well as for a topological
generalization of the PSM associated to $H$-Poisson
manifolds)\footnote{This corrects an inaccurate statement in
\cite{G/G}, where only invariance of $B$ under the Hamiltonian vector
fields generated by $\CP$ was required.  Likewise, if one adds a
WZW-like $H$-term: invariance of the closed 3-form $H$ would require
the contraction (now a 2-form) to be closed only (instead of to be
zero as turns out to be necessary for gauge invariance).}  \be
\int_\Sigma \CX^\ast B \, . \label{B} \ee Likewise the local
symmetries are not spoiled, if one adds e.g.\ a term of the form \be
\int_{\S} C(X(x)) \varepsilon \, , \label{nontop} \ee where $C$ is a
Casimir function of $\CP$ and $\varepsilon$ a 2-form on $\S$. (More
generally, one may also add a sum of such terms, with several Casimir
functions and volume forms).  Due to the appearance of the 2-form(s)
on $\S$, the action is no longer topological in this case. Still,
several features of the theory, including the structure of the moduli
space of classical solutions considered in this paper, remain
unaltered; consequently, the theories are called almost topological.

Poisson Sigma Models are of interest for at least three
reasons. First, they provide a unifying framework for several
two-dimensional field theories \cite{PSM1,G/G,PSM3,TK1,Habil1},
including gravity and Yang-Mills gauge theories. Within this paper we
will use part of this relation to check more general considerations. 

The second point of interest in the topological Sigma Model
\re{action}) stems from its significance for the quantization of
Poisson manifolds. It was noticed already early on  \cite{PSM2,G/G}
(cf also \cite{StMarg}) that
the quantization of the two-dimensional field theory (\ref{action}) is
closely related to the quantization of the target manifold,
interpreted as the collection of phase spaces for fictitious point
particles (namely the symplectic leaves). Indeed, in a Hamiltonian
quantization, $\Sigma$ taken cylindrical, any physical quantum state
of the theory corresponds to a symplectic leaf $(L,\Omega)$ satisfying
the integrality condition\footnote{In fact it is true in this form
only if the respective symplectic leaf $L$ is simply
connected. Otherwise there are additional states, in general also
corresponding to nonintegral leaves. Cf \cite{Habil1} for a complete
set of conditions.}  \be \oint_\sigma \Omega = 2\pi n \hbar \, , \quad
n \in \dN \; , \qquad \forall \sigma \in H^2(L) \, . \label{quant} \ee
This relation is readily recognized as the condition for
quantizability of the respective symplectic leaf $L$ in the framework
of Geometric Quantization (cf e.g.\ \cite{Woodhouse}).

An approach to quantization applicable to general Poisson manifolds
$M$ is provided by the program of Deformation Quantization (cf e.g.\
\cite{BayenI,BayenII}). Here the main idea is to find an {\em
associative\/} (local) deformation of the product of functions on $M$
in the form of a formal power series which in next to leading order in
the deformation parameter coincides with the Poisson bracket. A
solution to this by then already long-standing mathematical problem
was provided by Kontsevich \cite{Kontsevich}. Kontsevich's formula,
finally, received an illuminating interpretation \cite{CF1} as
appropriate two-point correlation function (evaluated on the boundary
of a disc, $\Sigma\approx\dR^2$) in a perturbation expansion of
\re{action}).

Last but not least, a Poisson Sigma Model may be regarded as an
appropriate zero slope limit of (perturbative) String Theory in the
background of a $B$-field. This may be a particularly interesting
point of view in the context of open strings ending on D-branes, where
in the case of a {\em constant\/} background, the effective Yang Mills
theory induced on the D-brane was seen to become noncommutative
\cite{Volker,SWmap}. The induced noncommutative product was found to
be the Moyal product \cite{Moyal} of the respective Poisson bivector
$\CP$ --- the antisymmetric part of the inverse of the sum of $B$ and
the closed string metric $g$ --- which is the (previously known)
specialization of the Kontsevich solution to the case of constant
$\CP$.  One of the open issues in this realm is the generalization of
this result to the case of nonconstant $B$-fields (resp.\ nonconstant
bivectors $\CP$). A reformulation of String Theory in terms of a
nontopological deformation of \re{action}) may provide the appropriate
link in this context. Such a relation shall be pursued elsewhere,
however.

In the present paper we focus on the moduli space $\Mod$ (denoted also
more explicitly by $\Mod(\S)$ or by $\Mod(\S,M,\CP)$) of {\em
classical\/} solutions of the (almost) topological models discussed
above. For a fixed topology of the worldsheet (two-dimensional
spacetime or base) manifold $\S$ and fixed target Poisson manifold
$(M,\CP)$, we are interested in all smooth solutions to the classical
field equations (stationary points of the action functional $S$),
where solutions differing only by a gauge transformation (specified
more clearly in section \ref{sec:generalities} below) are to be
identified.  For fairly reasonable topology of $\Sigma$ (guaranteeing
that $\rank H_1(\Sigma)$ is finite) this moduli space will be finite
dimensional.  We will not be able to find globally valid solutions for
completely arbitrary Poisson structures $(M,\CP)$, the main condition
being the existence of a presymplectic form $\wtO$ (on $M$ or at least
in a neighborhood of any symplectic leaf $L$ of $M$) which is
compatible with the Poisson bivector $\CP$, i.e.\ whose pull back to a
leaf $L$ coincides with the symplectic structure of $L$ as induced by
$\CP$ \cite{brackets}. This result can, however, also be used to
determine solutions if such a compatible presymplectic form only
exists for some subset of leaves in $(M,\CP)$. Moreover, information
about the space of solutions $\Mod(\Sigma,M,\CP)$ can still be found
in more general cases by gluing techniques.

If the boundary of $\Sigma$ is non-empty, there are some options as to
what kind of boundary conditions to place on the fields and
symmetries. In the present paper we allow for noncompact topologies of
$\Sigma$, but will not restrict fields or symmetries at the (ideal)
boundary. This may be motivated e.g.\ by physical models arising as
particular PSMs. At least for some mathematical applications (but also
e.g.~in gravitational models in the Hamiltonian formalism for ``open
spacetimes'', cf, e.g., \cite{Habil1}) it is, however, also of
interest to restrict some of the fields at the boundaries of $\Sigma$
and simultaneously to freeze all or some of the symmetries there; the
corresponding moduli space will in general differ from the above one
and shall be denoted by $\Mod^0$ (irrespective of the precise kind of
conditions at $\partial \Sigma$). For $\S = [0,1] \times \dR$ with
boundary conditions requiring that the pull-back of $A_i$ to
$\{0\}\times\dR$ and $\{1\}\times\dR$ vanishes and frozen symmetries
at these boundaries, $\Mod^0$ may be obtained also by symplectic
reduction, and for sufficiently well-behaved $(M,\CP)$ it was found in
\cite{CFHam} to carry the structure of a symplectic groupoid over $M$,
which integrates the Lie algebroid on $T^*M$ associated to the Poisson
manifold $M$. In this case the transition from $\Mod^0$ to $\CM$
essentially corresponds to an additional factorization, replacing $M$
by the corresponding leaf space induced by $\CP$.  Many of our results
can be adapted to the case of $\Mod^0(\S)$ for any $\S$, but we will
not do so explicitly in the present paper.

The organization of the paper is as follows. In Sec.\
\ref{sec:generalities} we discuss the field equations and symmetries
of Poisson Sigma Models.  We also briefly review how particular
choices of Poisson structures result in known theories such as
two-dimensional nonabelian gauge theories and the gravity theories
mentioned above. In the subsequent section, Sec.\
\ref{sec:othermethods}, we determine $\Mod(\S,M,\CP)$ for particular
choices of $(M,\CP)$ where methods other than the one presented in the
main part of the paper are available such that there are results to be
compared with. These are in particular the nonabelian gauge theories
as well as Poisson Sigma Models with a topologically trivial foliation
of the target manifold $M$, i.e.\ where $M$ is fibered by leaves of
trivial topology.

(Likewise results, not mentioned explicitely, are available for the
case of the general Sigma Model when one restricts to topologically
trivial worldsheet manifolds, $\S \approx \dR^2$. Furthermore, quite
explicit results exist for the case of the 2d gravity models
\cite{TK1,TK2,TK3,TKkinks}; due to an additional complication
resulting from a nondegeneracy condition to be satisfied by the
gravitationally acceptable solutions, these are mentioned only rather
briefly here and a more detailed analysis is deferred to later work.)

In Sec.\ \ref{sec:sols} we present our main results concerning the
solutions to the field equations of Poisson Sigma Models. They are
summarized in Theorem \ref{Summ}, providing the general solution for
Poisson structures permitting a compatible presymplectic form in a
neighborhood of any leaf, where the topology of $\S$ is arbitrary. We
are also able to integrate the symmetries effectively,
Eqs.~\re{symmetries}) below, such that we can provide representatives
of any gauge equivalence class of the space of solutions. Leaves which
do not have a neighborhood permitting a compatible presymplectic form
are not covered by the theorem, but we provide a discussion of a
possible generalization. (Let us mention right
away that there are Poisson manifolds with prominent examples such as
the Lie Poisson manifold $\frg^*$, $\frg$ compact semisimple, where
there are {\em no\/} leaves permitting a compatible presymplectic
form.)

In Sec.~\ref{sec:ex} the results are specialized and compared to the
results of Sec.\ \ref{sec:othermethods}, in particular also to the
two-dimensional nonabelian $BF$-theories with noncompact Lie algebra
$\frg$. This example is also used in Sec.~\ref{sec:nonregular} to
illustrate the role of nonregular leaves. The analysis shows that in
the case of $BF$-theory nonregular leaves are related to the
irreducibility of connections. If the rank of the fundamental group of
$\S$ is at most one, nonregular leaves contribute only a subset of
lower dimension to the solution space $\Mod(\S)$. For sufficiently
large rank, the highest dimensional stratum of $\Mod(\S)$ is obtained
for irreducible connections corresponding to the origin as a
nonregular leaf. The methods of this paper then give information on
lower dimensional subsets of the solution space $\CM$ with reducible
connections. For Yang--Mills theories these are solutions with
non-vanishing electric field, whereas an irreducible connection can
only exist when the electric field vanishes. The results for
Yang--Mills theories together with those of other methods indicate
that the classification of solutions can be generalized
straightforwardly to non-regular leaves, even though the explicit
formula for the solution $A$ in Theorem \ref{Summ} does not hold true.

\section{Setup}
\label{sec:generalities}

\subsection{Field equations, gauge symmetries and moduli spaces}
\label{sec:eom}

For a chosen local coordinate system in $M$, the fields can be
understood as a collection of scalar fields $X^i$ and 1-forms $A_i$,
$i=1,\ldots,n$, $n\equiv \dim M$, living on $\S$.  In general, this
works on local patches of $\Sigma$ only, however; still this
perspective is sufficient for most of the purposes of the present
paper.  These patches on $\Sigma$ may still be larger than those where
local coordinates $x^\mu$, $\mu = 1,2$, on $\Sigma$ exist and in which
$A_i = A_{i\mu}(x) \md x^\m$. (E.g.\ if $M$ is chosen as the dual of a
Lie algebra, $X^i$ may be taken as linear coordinates on $M$. Then
$X^i$ and the 1-forms $A_i$ exist globally on $\Sigma$, irrespective
of the topology of the worldsheet manifold).  For some purposes it is
also convenient to consider $A_i$ as a Grassmann odd field on
$\Sigma$; in this context (such as in the field equations below)
functional derivatives are understood as left derivatives always and,
unless otherwise stated, products of forms (Grassmann objects) are
understood to be wedge products.

The field equations of the action functional (\ref{action}) are
\ba \frac{\delta S}{\delta A_i} \equiv \md X^i + \CP^{ij}(X) A_j = 0
\label{eqs1} \\   \frac{\delta S}{\delta X^i} \equiv \md A_i + \2
\CP^{kl}{},_{i}(X) \, A_k A_l = 0 \label{eqs2} \ea If the terms
(\ref{B}) and (\ref{nontop}) are added only the second of these
equations changes, since both terms depend on the field $X$ only.
Moreover, the contribution of (\ref{B}) to the second equations, $\2
H_{ijk} \md X^j \md X^k$, vanishes upon use of (\ref{eqs1}) and the
condition imposed on $H = \md B$! So, the addition of (\ref{B}) or a
similar WZW-term has neither an effect on the field equations nor on
the local symmetries (cf below) and consequently the classical moduli
space is unchanged. A more interesting modification of the topological
PSM is obtained when one drops the condition of a vanishing
contraction of $H$ with the bivector while simultaneously replacing
the Jacobi identity for the Poisson bracket by the more general
condition $\CP^{ij}{},_{s} \CP^{sk} + \mbox{cycl}(ijk) = \CP^{ir}
\CP^{js} \CP^{kt} H_{rst}$ characterizing an H-Poisson structure (cf
\cite{Ctirad,Park,3Poisson}); this changes both field equations and
symmetries, but will not be further pursued in the present paper. The
contribution from (\ref{nontop}) to the lefthand side of (\ref{eqs2}),
on the other hand, is simply $C,_{i} \varepsilon$ or, more generally,
$C^\s{},_{i} \varepsilon_\s$, where several Casimir functions
$C^\s$ and 2-forms $\varepsilon_\s$ have been introduced, which
explicitly breaks the topological nature of the equations:
\begin{equation}
  \md A_i+\2\CP^{lm}{},_{i}(X) A_l A_m +C^\s{},_{i}(X)\varepsilon_\s
  = 0\,. \label{eqsnontop}
\end{equation}

We remark in parenthesis that the field equations (\ref{eqs2}) are
covariant with respect to target space diffeomorphism induced changes
of field variables only if the field equations (\ref{eqs1}) are
used. Given some auxiliary connection $\Gamma^i{}_{jk}$ on $M$, this
may be cured by replacing (\ref{eqs2}) by \be \mbox{D} A_i + \2
\CP^{kl}{}_{;i}(X) \, A_k A_l = 0 , \ee where $\mbox{D} A_i = \md A_i
- \Gamma^j{}_{ik} \md X^k A_j$ ($D$ is the induced exterior covariant
derivative acting on forms taking values in the pullback bundle $\CX^*
T^*M$) and the semicolon denotes covariant differentiation
with respect to $\Gamma$.

Next we turn to the symmetries of the action (\ref{action}). It is
straightforward to check that under the infinitesimal symmetry
transformations
\be \delta_\epsilon X^i = \epsilon_j \CP^{ji}(X) \; , \; 
\delta_\epsilon A_i = \md \epsilon_i + \CP^{kl}{},_{i}(X) A_k \epsilon_l
\label{symmetries}\; , \ee
where the $\epsilon_i$ are arbitrary functions on $\Sigma$, 
the action changes only by a total divergence $\int_{\Sigma}
\md(\epsilon_i\md X^i)$ thanks to the Jacobi identity
\be \CP^{il}\CP^{jk}{},_{l}+ \CP^{jl}\CP^{ki}{},_{l}+
\CP^{kl}\CP^{ij}{},_{l}=0 \label{Jacobi} \ee
for the Poisson tensor. Almost topological
models have the same symmetries due 
to the definition of a Casimir. 

The set of gauge transformations (\ref{symmetries}) is (in general
slightly over-)complete, or, in other words, it is an (in general
slightly reducible) generating set of gauge transformations (cf
\cite{HT} for definitions and further details).\footnote{At least for
topologies of $\Sigma$ with $\rank H_1(\Sigma) \le 1$ this is obvious
from a Hamiltonian analysis of the theory, cf e.g.~\cite{PSM1}. For
general topologies of the worldsheet it may, strictly speaking,
require a separate proof. In any case, we will consider as symmetries
to be factored out all those that are generated by
(\ref{symmetries}).}  This in particular implies that {\em any\/}
other local (gauge) symmetry of (\ref{action}) (invariance of the
functional, parametrized by some set of arbitrary functions on
$\Sigma$)  can be expressed in terms of
(\ref{symmetries}) up to so called trivial gauge transformations {\em
and\/} with possibly field dependent parameters $\epsilon_i$. If
$y^\alpha$ denotes the set of all fields of the action functional $S$,
$S=S[y^\alpha]$, $\alpha$ being a collective index, then trivial gauge
transformations are (infinitesimally) of the form $\delta_\mu y^\a =
\mu^{\a \b} (\delta S/\delta y^\b)$ (the sum also involving an
integration) for some graded ``antisymmetric'' but otherwise arbitrary
$\mu$. They are called trivial because on-shell (that is on the space
of solutions to the field equations) they act trivially and because
they exist for any action functional $S$. According to Theorem 17.3 of
\cite{HT} {\em any\/} symmetry of (\ref{action}) vanishing on-shell is
of this form.

The trivial transformations form a normal subgroup $\CN$ of all the
gauge transformations $\bar \CG$. It is only the respective quotient 
group $\CG = \bar \CG/\CN$ that is of relevance for the
gauge identification of solutions. The infinitesimal gauge
transformations certainly form a Lie algebra (the infinite-dimensional
Lie algebra of $\bar \CG$). However, the representatives
(\ref{symmetries}) do not; instead one finds: \be [\delta_\epsilon,
\delta_{\tilde \epsilon}] = \delta_{[ \epsilon , \tilde
\epsilon]} + \int_\Sigma \epsilon_j \tilde \epsilon_i
\CP^{ij}{},_{kl} \frac{\delta S}{\delta A_k} \frac{\delta }{\delta
A_l} \; , \ee where $[ \epsilon , \tilde \epsilon]_k \equiv
\epsilon_i \tilde \epsilon_j \CP^{ij}{},_{k}(X)$.  
 This is of the expected form since the commutator of two
gauge transformations is another one, and the representatives
(\ref{symmetries}) are complete. Also the (field dependent)
coefficient in front of the contribution vanishing on-shell is indeed
symmetric in the Grassmann part. (But even in the absence of the
latter contribution, making the algebra of (\ref{symmetries}) an
``open'' one, the field dependence of the new parameter
$[ \epsilon , \tilde \epsilon]_i$ spoils the Lie
algebra property.)  Completeness ensures also that the obvious
worldsheet diffeomorphism invariance of (\ref{action}) can be
expressed in terms of (\ref{symmetries}); indeed for any generating
vector field $\xi \in \Gamma(T\Sigma)$ one finds for the respective
Lie derivative acting on the space of fields under consideration \be
\CL_\xi = \delta_{\langle A , \xi \rangle} + \int_\Sigma \langle
\frac{\delta S}{\delta A_i} , \xi \rangle \frac{\delta }{\delta X^i} -
\int_\Sigma \langle \frac{\delta S}{\delta X^i} , \xi \rangle
\frac{\delta }{\delta A_i} \approx \delta_{\langle A , \xi \rangle} \;
,
\label{diffeo} 
\ee where the latter weak equality sign $\approx$ is used to denote on-shell
equality. 

The moduli space $\Mod$ is now defined as the space of all gauge
inequivalent smooth fields $X^i(x)$ and $A_{i\mu}(x) \md x^\mu$ on a
fixed worldsheet $\Sigma$, which we denote collectively by $\Phi$,
satisfying the field equations (\ref{eqs1}) and (\ref{eqs2}): \be \Mod(\S) =
\frac{\left\{\Phi \, | \; \delta S/\delta \Phi =
0\right\}}{\mbox{gauge equivalence}} \; , \label{mod} \ee where
``gauge equivalence'' is the equivalence relation generated by
(\ref{symmetries}); in particular the gauge group, called $\CG$ above,
is taken to be connected and simply connected. According to
(\ref{diffeo}), on the space of solutions infinitesimal
diffeomorphsims are generated by the symmetries under consideration;
correspondingly, {\em at least\/} the component of unity of
$\Diff(\Sigma)$ will be factored out in (\ref{mod}). It may also
happen, however, that several disconnected components of
$\Diff(\Sigma)$ fit into $\CG$, and thus are factored out in
$\Mod$---\cite{PLB} provides an explicit example for this scenario.

Note that in the general case the gauge symmetries are known explicitly 
only in their infinitesimal form \re{symmetries}). To determine whether 
two given solutions are gauge equivalent, we will, however, have to 
integrate these symmetries in one way or another. 

According to Noether's second theorem
the existence of
nontrivial gauge symmetries implies dependencies among the field
equations. With (\ref{symmetries}) one finds \be \CP^{ij}(X) \,
\frac{\delta S}{\delta X^j} \equiv \CP^{ij}{},_{k}(X) A_j \frac{\delta
S}{\delta A_k} + \md \frac{\delta S}{\delta A_i} \, .
\label{noether}  
\ee 
A discussion of the relevance of Noether's identities in the BV
formalism with a special emphasis on the example of the PSM may be
found in \cite{Stasheff}. 

As mentioned already in the Introduction, in the case that $\Sigma$
has (ideal) boundary components, one possibly wants to impose some
additinal boundary conditions on the admissible fields $\Phi$ and on
the parameters $\epsilon_i$ in (\ref{symmetries}). Acceptable boundary
conditions on $\Phi$ should be such that $\int_{\partial \S} A_i
\delta X^i$ vanishes, so that (\ref{action}) has well-defined
functional derivatives.  This is guaranteed e.g.~by the
Cattaneo-Felder boundary conditions $\langle A_i , v \rangle (x) = 0$
for any $x \in \partial \S$ and $v \in T_x \partial \S \subset T_x
\S$.  Alternatively, one may also
consider some Dirichlet-type boundary conditions on $\CX$ (more
generally, some mixture of both types of conditions, cf
e.g.~\cite{Habil1}), where the image of $\partial \S$ has to
lie within one symplectic leaf of $(M,\CP)$ so as to permit nontrivial
solutions of the field equations.

Note that the moduli space (\ref{mod}) does not change if the
symmetries are restricted {\em correspondingly\/} on the boundary,
i.e.~if any gauge orbit of the unrestricted theory leading to the
moduli space $\Mod$ is in 1-1 correspondence with a gauge orbit of the
restricted gauge equivalence on the set of the restricted
fields. E.g.~in the particular model (\ref{action}) with $\CP \equiv
0$ and upon the choice of the CF boundary conditions this would imply
that $\epsilon_i$ should be constant along the boundary. Requiring,
instead, $\epsilon_i$ to vanish on all of $\partial \S$ then enlarges
the moduli space to a bigger one, $\Mod^0$, which yields $\Mod$ only
after taking another  quotient. 

In the gravitational context there will be also other moduli spaces of 
relevance, denoted collectively by $\Mod^{\mathrm{grav}}(\S)$; we will 
discuss this briefly below.

As alluded to already above, not all of the symmetry generators
(\ref{symmetries}) are independent, at least if the topology of
$\Sigma$ is nontrivial and $\CP$ has a nontrivial kernel. For
illustration let us consider a Poisson tensor for which the first $k$
coordinates $X^i$, $i=1, \ldots, k$, are Casimir functions.  Then, for
any choice of $\epsilon_i$ nonvanishing only in the first $k$ indices,
one has $\delta_\epsilon X^i =0$ and $ \int_\alpha \delta_\epsilon A_i
=\int_\alpha \md \epsilon_i= 0$ for any closed loop $\alpha$. This
corresponds to $k$ times $\rank H^1(\S)$ nontrivial (global) relations
between the generators (\ref{symmetries}). In a BRST or BV
quantization scheme this requires ghosts for ghosts (for an explicit
construction of the respective Hamiltonian BRST charge for $\S = S^1
\times \dR$ cf.~\cite{PSM1}). This complication is e.g.~absent if $\S$
is a disc as in \cite{CF1} since $H^1$ vanishes in this case.

\subsection{Special cases}
\label{sec:special}

As mentioned already in the Introduction, there are several particular
cases of models of the type \re{action}) for which $\Mod(\S)$ (or at
least some facts about $\Mod(\S)$ like its dimension) is known
or can be determined by other means. 

\subsubsection{Two-dimensional nonabelian gauge theories}
\label{NonAb}

The most obvious of these is the specification to a nonabelian gauge
theory resulting from a choice of $(\CP^{ij})$ linear in $X$,
$\CP^{ij} = f^{ij}{}_k \, X^k$.  In this way, $M\equiv\frg^*$ can be
identified with the dual of the Lie algebra defined by the structure
constants $f^{ij}{}_k$, equipped with the Kirillov--Kostant Poisson
structure defined by $\CP_X(\alpha,\beta)=X([\alpha,\beta])$ where
$X\in\frg^*$ and $\alpha,\beta\in T_X^*M\equiv\frg^{**}\equiv\frg$. In
coordinates $X=X^iT_i$ with generators $T_i$ of $\frg^*$, this yields
the above linear Poisson tensor. In what follows, we will restrict
ourselves to semisimple Lie algebras such that the dual of $\frg$ can
be identified with $\frg$ itself by means of the Cartan--Killing
metric, which we will denote by $\tr$ below. We then express the
fields as the Lie algebra valued functions $A=A_iT^i$ and $X=X^i T_i$
on $\S$ for which the field equations \re{eqs1}), \re{eqs2}) may be
rewritten as $F\equiv \md A + A \wedge A =0$ and $D_A \, X \equiv \md
X + [A,X] = 0$. Furthermore, the gauge symmetries \re{symmetries}) can
be integrated to the equivalence relations $A \sim A^g \equiv g^{-1} A
g + g^{-1} \md g$, $X \sim X^g \equiv g^{-1} X g$ where $g(x)$ is an
arbitrary $G$-valued function on $\S$.  We restrict ourselves to
trivial bundles $\Sigma\times G$; for nontrivial bundles, already
$\CX$ is no more just a map $\Sigma \to M$, and much of what has been
said above needs reformulation (cf \cite{PSM1,bks} for such an
attempt).

The gauge transformations deserve further mention: $g(x)$ denotes a
map from the worldsheet $\S$ to the chosen structure group $G$ whose
Lie algebra $\frg$ has structure constants $f^{ij}{}_k$. The choice of
$G$ is not unique, different choices differing by the fundamental
group $\pi_1(G)$. A nontrivial $\pi_1(G)$, on the other hand, gives
rise to ``large'' gauge transformations for a multiply connected
worldsheet $\S$, i.e.\ gauge transformations not being connected to
the identity and thus not resulting from a direct integration of the
infinitesimal form of the symmetries. Thus, for a better comparison,
$G$ should always be taken to be the uniquely determined simply
connected group having the given structure constants
$f^{ij}{}_k$.\footnote{In some cases, as e.g.\ for the Lie algebra
$sl(2,\dR)$, this excludes the possibility of a matrix representation
of $G$. It is, however, still possible to construct the moduli space
along the lines below.}

Yang--Mills theories are obtained by adding the part \re{nontop}) with
the quadratic Casimir $C(X)=\tr(X^2)$ and a volume form $\varepsilon$ on
$\Sigma$ to the action. The field equation for $X$ is unchanged while
the zero curvature condition turns into
$F=-C,_{i}T^i\varepsilon=
-X\varepsilon$, rendering $X$ to play the role of the electric field. 
 As in the general case, gauge transformations are not
changed by adding \re{nontop}).

\subsubsection{Two-dimensional gravity models}
\label{s:2dgrav}

Let us choose $M \simeq \dR^3$ with coordinates $(X^i) := (X^+, X^-,
\phi) \in
\dR^3$ and a Poisson bracket defined through $\{X^+,X^-\}= W(\phi)$,
$\{X^\pm, \phi\} = \pm X^\pm$ for a (sufficiently) smooth function
$W$.  Such a bracket has one Casimir function $C$ given by\footnote{To
fully define $C$ by this equation, one may choose some fixed constant
for the lower bound of the integral.}
\be C = 2X^+X^- - 2 \int^\phi W(z) \md z \; . \label{Cas} \ee
Interpreting $X^a$, $a \in \{+,-\}$, as a Lorentz vector in a
two-dimensional Minkowski space, the bracket $\{ \cdot, \cdot \}$ is
seen to be invariant with respect to the corresponding Lorentz
transformations. Actually, for this purpose and most of what is
described below, $W$ could be allowed also to depend on the Lorentz
invariant combination $X^+X^-$; for simplicity, however, we restrict
ourselves to functions of the third coordinate $\phi$ only.

Identifying the respective 1-forms $A_\pm$ with a zweibein $e_\pm =
e^\mp$ (using the Minkowski metric of a frame bundle to raise and
lower indices) and $A_\phi$ with a spin connection 1-form $\omega$
($\omega^a{}_b = \varepsilon^a{}_b \, \omega$, $\varepsilon$
antisymmetric and normalized according to $\varepsilon^{+-} = 1$), and
upon dropping a surface term, the action \re{action}) assumes the form
\be S^{\rm grav} = \int_\S \,\, X_a De^a + \phi \md\o + W(\phi) \,
\varepsilon \,\, , \label{grav} \ee where $De^a \equiv \md e^a +
\o^a{}_b \wedge e^b$ is the torsion 2-form and $\varepsilon = e_+
\wedge e_-$ is the (\emph{dynamical}) volume form on $\S$. Thus $S$ is seen
to yield an action for a gravitational theory defined on a
two-dimensional spacetime $\S$.

In fact, e.g.\ if $W$ is a convex function, we may eliminate the
fields $X^i$ by means of their algebraic field equations, and the
action may be seen to take the purely geometrical form \be S^{\rm
geom}[g] = \int_\S \md^2x \sqrt{-g} f(R) \, , \label{fR} \ee where $f$
is the Legendre transform of $-2 W$ and $R$ is the Ricci scalar of the
torsion free Levi Civita connection associated to the two-dimensional
metric $g_{\mu\nu}=2e_\m{}^+ e_\n{}^-$. The prototype of such a higher
derivative theory is provided by $R^2$ gravity, $f(R)=\frac{1}{8} R^2
+ 2$, resulting from the quadratic Poisson bracket defined by $W=1 -
\phi^2$.

There are also more general possibilities to obtain gravitational models 
as particular PSMs. We will come back to this elsewhere \cite{prep}. 

\section{Results by other methods}
\label{sec:othermethods}

In special cases of the Poisson manifold $(M,\CP)$, there are known
methods to solve the field equations which will be recalled here for
later use.

\subsection{$\Mod(\S)$ for topologically trivial Poisson manifolds}
\label{Trivial}

Let us assume within this subsection that the Poisson manifold
$(M,\CP)$ is foliated regularly into symplectic leaves and that the
typical leaf admits a set of {\em globally\/} defined Darboux
coordinates $X^\a$.  Indexing the leaves by (possibly only locally
defined) coordinates $X^I$, $I=1,\ldots,k$, (any set of independent
Casimir functions on $M$ where $k$ is the dimension of the kernel of
$\CP$, which, by assumption, is constant all over $M$), there thus
exists a combined coordinate system $(X^i)=(X^I,X^\a)$ on $M$ for
which the matrix $\CP^{ij}$ is zero everywhere except for the block
$\CP^{\a \b}$ which has standard Darboux form.

In such a coordinate system the field equations \re{eqs1}), \re{eqs2})
are trivial to solve: Clearly, the first set of equations implies that
the $X^I(x)$ are constant functions all over $\S$
and that the $A_\a$ are determined uniquely by means of $X^\a(x)$; if
$\O_{\a\b}$ denotes the (constant) inverse to $\CP^{\a\b}$, $A_\a =
\O_{\b\a} \md X^\b$.  The nontrivial content of the second set of the
equations, furthermore, reduces to $\md A_I=0$. Also the local
symmetries \re{symmetries}) can be integrated easily in this case. One
learns that any smooth choice of the set of functions $X^\a(x)$ is
gauge equivalent to any other choice and that any $A_I$ just
transforms like a $U(1)$ gauge field, i.e.\ $A_I \sim A_I + \md h_I$
for any (smooth) function $h_I(x)$.

Thus, restricting $M$ further to $\dR^n$ for simplicity, we obtain
\begin{prop}
  Let $(\dR^n,\CP)$ be a trivially foliated Poisson manifold with
  leaves homeomorphic to $\dR^{n-k}$ where $k=\dim\Kern\CP$.  A set of
  representatives of the gauge equivalence classes of solutions to the
  field equations is given by \be X^I(x) = C^I = const. \; , \; \:
  \quad X^\a(x) = 0 = A_\a \; , \quad \; \: A_I = \a_I \; \: \;
  \mbox{with}\; \; [\a_I] \in H^1(\S) \, .  \ee Correspondingly, the
  classical moduli space is found to be of the form: \be
  \label{dimM}\Mod(\S) = \dR^{k(r+1)} \qquad \mbox{with} \quad \; r
  \equiv \rank H^1(\S) = \rank \pi_1(\S) \quad \mbox{and} \quad k =
  \dim \Kern \CP \; .\ee
\end{prop}

Here, $k$ is also the (by assumption constant) codimension of the
symplectic leaves in $M$. As usual, $H^1(\S)$ denotes the (always
abelian) first cohomology of the base manifold $\S$, its rank being
the number of independent generators.

Note that in the original formulation of the action \re{action})
admitting the above mentioned coordinates on $M$, as obtained, e.g.,
in the case of a gravity model, the matrix $\CP$ as a function on $M$
may be highly nonlinear. Then the change of coordinates implicitly
used above provides a trivialization of the otherwise possibly not
that simple field equations (for a partial illustration of this point
cf \cite{TK1}).

In the above we assumed that the symplectic foliation of $M$ is
topologically trivial while $\S$ was kept arbitrary. Alternatively,
one may achieve similar results (cf, e.g., \cite{PSM1,TK1}) for the
case that one restricts attention to a
local, topologically trivial patch of $\S$. The reason for this is
that any (generic) point in $M$ has a neighborhood which admits
Casimir--Darboux coordinates putting $\CP$ in the above mentioned
standard form \cite{WeinsteinDarboux}. It is the intention of the
present paper to extend these results to the case where $\S$ is left
completely arbitrary, but now $M$ need not necessarily be foliated by
symplectic leaves homeomorphic to a linear space (as implied by the
global existence of Darboux coordinates). E.g., we will be able to
cover cases such as those where $M$ is (regularly) foliated by
topologically nontrivial leaves provided only that they have trivial
second cohomology. For specialization to topologically trivial leaves
the above results will then be regained.

\subsection{Gauge theories}
\label{NonAbSol}

It is precisely in the case of linear Poisson structures in the
setting of Sec.\ \ref{NonAb} that the second set of field equations
\re{eqs2}) decouples from the first set \re{eqs1}), being a set of
equations for $A$ only. The moduli space of solutions to $F=0$ (the
space ${\cal A}_0$ of flat connections) modulo gauge symmetries is
well known: a flat connection is characterized uniquely by its
holonomies (elements of $G$) along a set of representatives for
generators of $\pi_1(\S)$, and gauge transformations act thereon by
joint conjugation. Therefore, the space of flat connections modulo
gauge transformations is given by ${\rm Hom}(\pi_1(\S),G)/\Ad_G$ where
$\Ad_G$ denotes the adjoint action of $G$, and any gauge invariant
function on the space of flat connections can be written as a function
of traces of holonomies, called Wilson loops, along generators
of the fundamental group. Note that for a compact structure group $G$
the space of flat connections modulo gauge transformations is compact.
This space is in general not a manifold but only a stratified space
because the adjoint action may have fixed points on ${\cal A}_0$. The
smooth part of $\Mod(\S)$ is given by the space of gauge equivalence
classes of those flat connections which yield holonomies with
centralizer in $G^r$, $r\equiv {\rank\pi_1(\S)}$, of minimum dimension. If
$r$ is large enough, these are irreducible connections; the latter 
can be defined as connections for which there is no non-central
element $g\in G\backslash Z(G)$ commuting with all holonomies.

Given a flat connection $A$, one can determine $X$ by solving the
equation $\md X=-[A,X]$, which locally can be integrated to
$X(x)=\Ad_{g_{x_0,x}} X(x_0)$ where $x_0$ is an arbitrary fixed point
in $\S$ and $g_{x_0,x}$ denotes the parallel transport between $x_0$
and $x$ along an arbitrary path. In a simply connected neighborhood of
$x_0$, the parallel transport $g_{x_0,x}$ is unique (i.e., independent
of the path) because $A$ is flat, and so $X$ is well defined. If $\S$
is not simply connected, the initial value $X(x_0)$ has to fulfill $r
\equiv \rank\pi_1(\S)$ integrability conditions because
$X(x_0)=\Ad_{h_{x_0}} X(x_0)$ for any holonomy $h_{x_0}$ along a
closed curve based in $x_0$. Given an irreducible flat connection $A$,
evidently these conditions cannot be fulfilled non-trivially for a
semisimple Lie group $G$, $X \equiv 0$ being the only admissible
solution in this case.

Only for a reducible flat connection $A$ can there be non-trivial
solutions $X$.  The explicit form, in some gauge, is determined by a
given connection $A$ via the field equation $\md X=-[A,X]$ , which
infinitesimally expresses the fact that a change in $X$ is given by
conjugation. Therefore, the image of $\S$ under the map
$\CX\colon\S\to\frg$ corresponding to $X$ has to lie entirely within
an adjoint orbit of $\frg$, the explicit form being determined by a
particular flat connection. If only ``small'' gauge transformations
are allowed, gauge transformations acting on $X$ correspond to smooth
deformations of the map $\CX$ within an adjoint orbit, demonstrating
that gauge equivalence classes of solutions $X$ are given by
particular homotopy classes of maps from $\S$ to some adjoint orbit in
$\frg$.

Given a semisimple Lie Group $G$ of rank $k$, there are always
irreducible flat connections on the Riemann surface $\S$ {\em
provided\/} that $\rank\pi_1(\S)$ is large enough. If $\Sigma$ has
genus $g$ and $n$ holes (boundary components), the fundamental group
can be represented by $2g+n$ generators $a_1,\ldots, a_g$,
$b_1,\ldots,b_g$ and $m_1,\ldots,m_n$ with one relation
$[a_1,b_1]\cdots [a_g,b_g]m_1\cdots m_n=1$ using the commutator
$[a,b]:=aba^{-1}b^{-1}$. A flat connection is uniquely specified by
its $2g+n$ holonomies in $G$ around the generators, which also have to
be subject to the given relation (eliminating one free holonomy).
Furthermore, factoring out gauge transformations generically
eliminates $\dim G$ parameters for choosing the holonomies. Thus, we
have a maximal dimension $(\rank\pi_1(\S)-2)\dim G=(2g+n-2)\dim G$ for
the space of irreducible flat connections modulo gauge transformations
(see also \cite{Goldman}). This shows that $\rank\pi_1$ has to be
larger than two in order to allow irreducible connections.

Thus, in general the moduli space $\Mod(\S)$ of gauge equivalence
classes of flat connections has a smooth stratum of maximal dimension
consisting of gauge equivalence classes of irreducible flat
connections (if $G$ is non-compact, this stratum may be non-Hausdorff,
but one can choose a dense subspace which is Hausdorff) of dimension
$(\rank\pi_1(\S)-2)\dim G$, whereas reducible flat connections give
lower-dimensional strata \cite{Goldman}.

Solutions for $X$ which lie in adjoint orbits of maximal dimension
(which is $\dim G-k$; this case corresponds to a flat connection with
holonomies generating a maximal abelian subgroup of $G$) contribute a
subset maximally of dimension $k(\rank\pi_1(\S)+1)$. The first
part, i.e.\ $k\rank\pi_1(\S)$, is the dimensionality of the space of
connections whose holonomies generate a maximal abelian subgroup of
$G$, whereas the contribution of dimension $k$ is the remaining
freedom in choosing $X$ (which is only free at a single point and has
to commute with all holonomies, i.e.\ it must also lie in the maximal
abelian subgroup of dimension $k$).  Clearly this is of lower
dimension if the rank of the fundamental group is large enough.  The
case of small $\rank\pi_1(\S)$ is special; in particular if
$\rank\pi_1(\S)\leq 1$, {\em all\/} flat connections are reducible and
a dense set in $\Mod(\S)$ is provided by connections which lead to
non-trivial $X$-solutions; this is also the case if $\pi_1(\S)$ is
abelian (e.g., if $\S$ is a torus).

For Yang--Mills theories, the curvature of $A$ is not necessarily zero
but given by $F=-X\epsilon$ in terms of the electric field
$X$. Therefore, the above strategy cannot be applied for a general
solution since holonomies are no longer invariant under deformations
of the curve defining $g_{x_0,x}$. This is the case only if $X$
vanishes where we have the same solutions as described above given by
flat connections. But in the physically more interesting case of
non-zero electric field $X$, a connection cannot be flat and solutions
for $A$ have to be determined by other means. This shows that
differences between $BF$- and Yang--Mills theories only arise in the
sector of non-vanishing $X$ which in $BF$-theories leads to reducible
flat connections and in Yang--Mills theories to non-flat
connections. The standard methods reviewed in the present subsection
start by using the mathematically well-studied space of irreducible
flat connections and are insensitive to those differences. On the
other hand, we will see that the methods of this paper are well-suited
to determine solutions with non-vanishing $X$-field and nicely
demonstrate the key difference between solutions to both theories.

\subsection{Gravity models}

In the gravitational setting, one is interested in maximally extended
solutions to the field equations resulting from a variation of the
action \re{grav}) for fixed topology of $\S$, having a globally smooth
and nondegenerate metric, and identifying solutions which are mapped
into one another by the gravitational symmetries, i.e.\ by
diffeomorphisms and local Lorentz transformations. This program has
been carried out in full generality in \cite{TK2,TK3,TKkinks},
yielding implicitly a description of the ``gravitational moduli
space'' $\Mod^{\rm grav}(\S)$, which, in particular, was found to be
finite dimensional for any fixed topology of $\S$.

With the methods of the present paper we will be able to derive
information about the space of solutions at a global level. Because
the requirement of a non-degenerate metric (including subtle relations
of the gravitational symmetries to the symmetries generated by
(\ref{symmetries})---cf also (\ref{diffeo}) and the explicit
discussion of this relation in \cite{PLB}) needs some care, however,
we will discuss this application elsewhere. Also other issues which
are specific to gravity, such as the completeness of the resulting
two-dimensional space-time, can be studied within the present setting.

\section{Moduli space of classical solutions}
\label{sec:sols}

In this section we present our results concerning solutions to the
field equations of Poisson Sigma Models. The topology of the
worldsheet $\S$ is taken to be fixed. Along possible boundary
components of $\S$ neither the fields nor the local symmetries will be
restricted in this context; for given $\S$ we look for the moduli
space of solutions to the field equations \re{eqs1}), \re{eqs2})
subject to the equivalence relation generated by the symmetries
\re{symmetries}).  The situation remains unchanged when boundary
conditions on the fields are added in the same ``number'' as
symmetries along the boundaries are frozen. In a Hamiltonian
formulation of the model on $\S \cong \dR \times \dR$, this is not
always adequate or in the line of a particular (physical or
mathematical) problem. Examples for this are two-dimensional gravity
models on open spacetimes as well as the recent considerations of
Cattaneo and Felder: In both cases additional parameters of the moduli
space appear by freezing {\em all\/} the symmetries on the boundary of
the first factor $\dR$ of $\S$ while only part of the fields are
subject to boundary conditions (in the case of Cattaneo and Felder
e.g.\ only the tangential components of $A_i$). The present method,
however, may be applied also to such cases.

\subsection{Solutions for $\CX$}

For the case that $\CP$ is linear, it was found in Section
\ref{sec:special} that it is advisable first to solve the equations
\re{eqs2}) for the $A$-fields, as (for the topological model) the
field equations for them decouple from the fields $X(x)$. In the
general case, however, the field equations \re{eqs2}) are much harder
to solve and in the present paper this will be achieved only under
certain conditions on the target manifold $(M,\CP)$. The field
equations \re{eqs1}) for the $X$-fields, on the other hand, can be
solved in full generality for arbitrary target, and even the local
symmetries may be integrated easily (although in a rather abstract
manner):

\begin{theo} \label{sol1} Let $(\CX\colon \S \to M,A)$ be a solution
to the field equations (\ref{eqs1}) and (\ref{eqs2}).

Then the image of $\CX$ lies entirely within one of the symplectic
leaves $L
\subset M$ of the foliation of $M$. All gauge equivalence classes
of solutions $\CX$ are provided by the homotopy classes of maps from
$\S$ to any $L$.
\end{theo}

\begin{proof} 
  Let $X\in\CX(\S)$ be a point in the image of $\CX$. We first assume
  that $X$ is a regular point of the foliation of $M$ into symplectic
  leaves, i.e.\ that $X$ lying in a symplectic leaf $L$ has a
  neighborhood $\CU$ homeomorphic to $(\CU\cap L)_X\times\dR^k$ with
  $k=\dim\Kern\CP(X)$ where $(\CU\cap L)_X$ denotes the connected
  component of $\CU\cap L$ containing $X$ (only in the case of a leaf
  $L$ which lies densely in a part of $M$ do we have $(\CU\cap
  L)_X\not=\CU\cap L$ for all $\CU$).  After choosing local
  coordinates in $\CU$ adapted to the decomposition into $\CU\cap L$
  and $\Kern\CP\cong\dR^k$, it is immediate to see that the components
  of $\CX$ along $\Kern\CP$ have to be constant in $\CU$ owing to Eq.\
  (\ref{eqs1}). Therefore, the image of $\CX$ lies in $L$ in a
  neighborhood of any regular point and the first assertion follows
  for the case of a regular foliation of $M$.
  
  In general, however, $M$ is not foliated regularly into symplectic
  leaves, meaning that there are also lower dimensional leaves which
  then lie in the boundary of a higher dimensional one. For $X$ lying
  in a lower dimensional leaf $L\subset\partial L'$, where $L'$ is a
  higher dimensional one, the above reasoning shows that all
  derivatives of components of adapted coordinates ``normal'' to $L$
  have to vanish. But this information alone is not sufficient to
  ensure that the image of $\CX$ lies entirely in $L$, for there are
  directions normal to $L$ but tangential to $L'$ leading to
  derivatives of $\CX$ which have to vanish only in $L$, not in
  $L'$. It is then possible to construct smooth maps $\CX\colon \S\to
  M$ which connect $L'$ with $L$. Using the complete field equations
  (\ref{eqs1}) for $\CX$, however, we can exclude such maps thanks to
  the uniqueness theorem for solutions of first order differential
  equations: Suppose there is a solution $\CX$ of (\ref{eqs1})
  connecting $L'$ with $L$ and a corresponding smooth solution $A$ of
  (\ref{eqs2}).  We can then find a smooth path $c$ of finite
  parameter length in $\S$ such that the interior of $\CX(c)$ lies in
  $L'$ and its endpoint in $L$. Eq.\ (\ref{eqs1}) then implies that
  the restriction of $\CX$ to $c$ is the integral curve of a smooth
  vector field (determined by $A$) vanishing at the endpoint of
  $\CX(c)$. The existence of such an integral curve reaching the
  singularity of the smooth vector field in a finite parameter
  distance is a contradiction.  This proves our first assertion in the
  general case.

  Using again local coordinates adapted to the foliation, it is
  easy to see that the gauge transformations (\ref{symmetries}) are
  infinitesimal homotopies of the map $\CX\colon \S\to M$, immediately
  leading to the second assertion.
\end{proof}

Since in the proof we only used the field equation \re{eqs1}) and
symmetries \re{symmetries}) for $X$ and the fact that $A$ is subject
to a first order differential equation which also holds true for an
almost topological model, we obtain the following

\begin{cor}\label{solnontop1}
 Let $(\CX\colon \S \to M,A)$ be a solution
 to the field equations (\ref{eqs1}) and (\ref{eqsnontop}).

 Then the image of $\CX$ lies entirely within one of the symplectic
 leaves $L \subset M$ of the foliation of $M$. All gauge
 equivalence classes of solutions $\CX$ are provided by the homotopy
 classes of maps from $\S$ to any $L$.
\end{cor}

Thus, although the model is no longer topological with the term
\re{nontop}), the $X$-solutions are still classified solely by
topological properties of the spaces $\Sigma$ and $M$ (the latter as a
foliated space).

\subsection{Compatible presymplectic forms}

The main tool for constructing solutions for $A$ corresponding to a
map $\CX$ into a leaf $L$ will be a presymplectic form $\widetilde
\O$, which, upon restriction to tangential vectors to any leaf $L'$ in
$U$ coincides with the respective symplectic 2-form $\O_{L'}$ induced
by the given Poisson bracket on $M$. Such a {\em compatible
presymplectic form\/} $\widetilde \O$ of $\CP$ does not exist under
all circumstances and if it exists, it will not be unique.

General conditions for the existence of such a 2-form $\widetilde
\O$, compatible with $\CP$ in the above sense, have been investigated
in \cite{VaismanBook} where the obstruction has been identified as the
characteristic form class of the Poisson bivector ${\cal P}$ and
recently in \cite{brackets} under the condition that there is an
integrable distribution transversal to the leaves in $U$, where they
have been put into the form of descent equations. Specializing these
equations to particular cases gives the following results which we
will use later:

\begin{cor}\label{triv} 
 If $M$ is foliated trivially, i.e.\ it is of the form $M\cong
 L\times\dR^k$, then a necessary condition for the existence of a
 compatible presymplectic form in a neighborhood of $L$ is
\[
 \partial_I\oint_{\sigma}\O_L=0 
\] 
 where $\partial_I$ denotes any differentiation transversal to $L$ and
 $\sigma$ is a closed 2-cycle in $L$. This means that the symplectic
 volume of any closed 2-cycle in a leaf has to be constant in $M$.
\end{cor}

The second result, which can be easily verified, applies to leaves of
trivial second cohomology:

\begin{lemma}\label{pot}
  If $\O_L$ has a symplectic potential $\theta_L$ on any leaf $L$ in
  $M\cong L\times\dR^k$, i.e.\ $\O_L=\mdp\theta_L$, and $\theta_L$
  varies smoothly from leaf to leaf, then $\wtO:=\md\theta$ is a
  compatible presymplectic form on $M$.

  In particular, if all leaves $L$ in a trivially foliated $M\cong
  L\times \dR^k$ have trivial second cohomology, then there exists a
  compatible presymplectic form on $M$.
\end{lemma}

The notation here is as follows: $\O_L$ and $\theta_L$ are
differential forms on the leaves which depend parametrically on
coordinates transversal to the leaves (e.g.\ Casimir functions). The
derivative operator $\mdp$ only acts on coordinates inside the leaf
$L$, whereas $\md$ is the exterior derivative in the embedding space
$M$ and acts on all coordinates; $\theta$, finally, by definition
coincides with $\theta_L$ on any leaf $L$. 

As already mentioned, a compatible presymplectic form is not
unique. Given one such form $\wtO$, one can always add a closed 2-form
$\lambda$ which vanishes when pulled back to the leaf, giving in fact
the complete freedom in defining $\wtO$
\cite{VaismanBook,brackets}. 

It is interesting to observe that conditions such as in Corollary
\ref{triv} have been found as obstructions to integrate the Lie
algebroid $T^*M$ to a smooth Lie groupoid, cf.~\cite{CFm1,CFm2}.

\subsection{Solutions for $A$}

We now may solve for the $A$-fields assuming a fixed map $\CX$ which
in particular singles out a symplectic leaf $L$.  (To obtain all
solutions, all possible leaves $L$ as well as representatives of all
homotopy classes from maps $\CX$ to $L$ have to be considered; in any
of these cases we then proceed by solving for $A$ for given $\CX$.)
In general, the field equation for $A$ is harder to solve, and we will
start our discussion with a special case and later discuss
generalizations.

\subsubsection{Solutions for topological models corresponding to
regular leaves of trivial holonomy}

We first assume that we are dealing with a map $\CX$ into a regular
leaf of trivial holonomy which allows a compatible presymplectic form
in a neighborhood.  By this assumption, we can choose a set of Casimir
functions $C^I$ in the neighborhood such that $L$ is given by the
preimage of zero (and $L$ has a neighborhood of the form
$L\times\dR^k$).  We then arrive at

\begin{prop} \label{sol2}
 For a given map $\CX$ with image in a symplectic leaf $L$ of trivial
 holonomy which has a neighborhood $U$ permitting a presymplectic form
 $\widetilde \O$ compatible with $\CP$, any solution to the field
 equations (\ref{eqs1}), (\ref{eqs2}) may be written in the form \be
 \label{solu}A_i = -\CX^\ast (\partial_i \lrcorner \widetilde \O ) +
 \a_I \; \CX^\ast (\partial_i C^I) \, , \ee where $C^I$,
 $I=1,\ldots,k$, are some Casimir functions with $L = (C^I)^{-1}(0)$
 and $\a_I$ are closed 1-forms on $\S$. For fixed $\widetilde \O$ and
 $C^I$, redefining $\a_I$ by adding an exact 1-form on $\S$ is a gauge
 transformation. For fixed map $\CX$, gauge equivalence classes of
 solutions to the field equations for $A$ correspond to the set of $k$
 elements $[\a_I] \in H^1(\S)$.
\end{prop}

Here, $H^1(\S)$ denotes the first cohomology of $\S$
over the real numbers. Eq.\ (\ref{solu}) demonstrates that, in
contrast to the moduli space of flat connections (see Sec.\
\ref{NonAb}), the $A$-solutions of Poisson Sigma Models corresponding
to $X$-solutions lying in regular leaves are not classified by the
fundamental group $\pi_1(\S)$ but by its abelianization, which is
$H^1(\S)$. This point will be clarified in the examples below when we
will reexamine the case of two-dimensional gauge theories from the
Poisson Sigma Model point of view.

\begin{proof} 
 Let $A$ be a solution to the equations (\ref{eqs1}) and (\ref{eqs2})
 and $\wtO$ be a presymplectic form compatible with $\CP$. We first
 introduce coordinates $(X^{\alpha},X^I)$ on the neighborhood $U$ of
 $L$ in $M$ adapted to the foliation such that the $X^{\alpha}$
 coordinatize a leaf $L$ and the $X^I$ are transversal, and show that
 in these coordinates $\underline{i}\in (\alpha,I)$
\begin{equation}\label{dA}
 \md(A_{\underline{i}}+\CX^\ast(\partial_{\underline{i}}\lrcorner\wtO))=0\,.  
\end{equation} 
 Using that $\wtO$ is compatible with $\CP$, Eq.\
 (\ref{eqs1}) immediately implies \begin{equation}\label{Aalpha}
 A_{\alpha}=-\wtO_{\alpha\beta}\md
 X^{\beta}=-\CX^{\ast}(\partial_{\alpha}\lrcorner\wtO) \end{equation}
 which shows (\ref{dA}) for tangential components.

 For transversal components, Eq.\ (\ref{eqs2}) with the expression for
 $A_{\alpha}$ leads to 
\[ 
 \md A_I+ \CP^{\alpha\beta}{},_{I}
 \wtO_{\gamma\alpha} \wtO_{\delta\beta} \md X^{\gamma}\md X^{\delta}=0
\] 
 where tangential components of the matrix $-\wtO\CP,_{I}\wtO$ appear.
 Owing to compatibility of $\wtO$, which for tangential components
 implies inverseness $(\CP\wtO)^\alpha_\beta= \delta^\alpha_\beta=
 (\wtO\CP)^\alpha_\beta$, as well as adaptedness of the coordinates,
 which implies $\CP^{\alpha I}=0=\CP^{IJ}$, the tangential components
 fulfill the equation $(\wtO\CP\wtO)_{\alpha\beta}=
 \wtO_{\alpha\beta}$. Taking a derivative with respect to $X^I$ yields
 $-(\wtO\CP,_{I}\wtO)_{\alpha\beta}= \wtO_{\alpha\beta,I}$ and thus
\[
 \md A_I+ \wtO_{\gamma\delta,I}\md X^{\gamma}\md X^{\delta}=0\,.  
\]
 Note that when we take the derivative, we need $\wtO$ to be compatible
 with $\CP$ in a whole neighborhood of the leaf and not just on the
 leaf itself. Using that the second term
 is nothing but the Lie derivative of $\wtO$ with respect to
 $\partial_I$ and that $\wtO$ is closed, we can reexpress this term as
 $\md\partial_I\lrcorner\wtO$. This proves Eq.\ (\ref{dA}) for all
 adapted coordinates.

 It now follows directly from \re{dA}) that
\[
 A_{\underline{i}}= -\CX^\ast(\partial_{\underline{i}}\lrcorner\wtO)+
 \alpha_{\underline{i}}
\] 
 for a set of closed 1-forms $\alpha_{\underline{i}}$ on
 $\S$. According to Eq.\ \re{Aalpha}) these 1-forms have to vanish for
 components of $A$ tangential to $L$. This then establishes Eq.\
 \re{solu}) as a necessary condition for the solutions
 $A_{\underline{i}}$ and also for $A_i$ in arbitrary coordinates since
 Eq.~(\ref{solu}) as well as Eqs.\ (\ref{eqs1}), (\ref{eqs2}) are
 target-space covariant.

  Sufficiency follows from the equivalence of Eqs.\ (\ref{eqs1}),
  (\ref{eqs2}) with (\ref{Aalpha}), (\ref{dA}) and the restriction on
  $X(x)$ found in Theorem \ref{sol1}.

  Eq.\ \re{solu}) is already covariant with respect to the gauge
  transformations of Theorem \ref{sol1} (a change of $X^\alpha(x)$
  induces the corresponding change of $A_\a$ according to this
  equation, which reduces to Eq.\ (\ref{Aalpha})). As seen best in
  adapted coordinates, independently of those transformations, the
  gauge transformations (\ref{symmetries}) allow us to change the
  transversal components $A_I$ by adding exact 1-forms (analogously to
  the discussion in Sec.\ \ref{Trivial}).  Thus, for a fixed map
  $\CX$, gauge equivalence classes of solutions are given by the
  cohomology classes $H^1(\S)$ of $\alpha_I$.  This demonstrates the
  last assertion of the Theorem.
\end{proof}

In the Proposition, the Casimir functions $C^I$ and the compatible
presymplectic form $\wtO$ were assumed to be fixed. Any other set
$C_2^I$ of Casimir functions can be obtained by a map
$C_2^I=f^I_J(C)C^J$ where $f^I_J$ are differentiable functions of the
original Casimir functions forming an invertible matrix. We then have
\[
 \CX^*(\partial_iC_2^I)=\CX^*(f^I_J\partial_iC^J+C^J\partial_if^I_J)=
 X^*((f^I_J+C^K\partial_Jf^I_K)
 \partial_iC^J)=f^I_J(0)\CX^*(\partial_iC^J)
\]
and the 1-forms $\alpha'_I$ corresponding to the Casimir functions
$C_2^I$ are given by 
\begin{equation}\label{transC}
 \alpha'_I=(f^{-1})_I^J(0)\alpha_J\,.
\end{equation}
This is just a linear recombination of the original 1-forms, but it
can still change the classes in $H^1(\S)$.

Choosing a different compatible presymplectic form $\wtO$ also implies
a redefinition of the 1-forms $\alpha_I$. As remarked after
Lemma~\ref{pot}, the freedom in $\wtO$ is given by adding a closed
2-form $\lambda$ which vanishes when pulled back to a leaf, i.e.\
$\md\lambda=0=\iota_L^*\lambda$. In a neighborhood of the leaf, such a
2-form can always be written as $\lambda=\beta_I\wedge\md C^I+
\gamma_{IJ}\md C^I\wedge\md C^J$ with 1-forms $\beta_I$ and functions
$\gamma_{IJ}$ which fulfill $\md\beta_I=-\partial_K\gamma_{IJ}\md
C^J\wedge\md C^K$. The last equation implies
$\md\CX^*\beta_I=\CX^*\md\beta_I=0$ since the image of $\CX$ lies in
the leaf $L$ where the $C^I$ are constant. If we change the compatible
presymplectic form to be $\wtO_2=\wtO+\lambda$, we obtain
\begin{eqnarray*}
 -\CX^*(\partial_i\lrcorner\wtO_2) &=& -\CX^*(\partial_i\lrcorner\wtO)-
  \CX^*(\partial_i\lrcorner\lambda)\\
 &=& -\CX^*(\partial_i\lrcorner\wtO)-\CX^*(\beta_I(\partial_i)\wedge\md
  C^I-\partial_iC^I\beta_I+2\partial_iC^I\gamma_{IJ}\md C^J)\\
 &=& -\CX^*(\partial_i\wtO)+\CX^*(\beta_I\partial_iC^I)
\end{eqnarray*}
and the 1-forms are changed to
\begin{equation}\label{transO}
 \alpha'_I=\alpha_I+\CX^*\beta_I\,.
\end{equation}
 As shown above, $\CX^*\beta_I$ is closed so that the new $\alpha'_I$
still define elements of the first cohomology. However, the
$\CX^*\beta_I$ need not be exact and so also a redefinition of $\wtO$
may change the cohomology classes characterising a given connection
(not just the representatives of the original classes).

This implies that there is no {\em canonical\/} isomorphism between
gauge equivalence classes of $A$-fields for fixed $\CX$ with the set
of $k$ elements in $H^1(\S)$, while still any particular choice of
$\widetilde \O$ and $C^I$ does define an isomorphism. The situation is
comparable to that in non-abelian gauge theories where a map between
gauge equivalence classes of gauge fields and elements of $\pi^1(\S)$
is defined only upon choosing closed curves which generate $\pi^1(\S)$.

\subsubsection{Generalizations}
\label{sec:Gen}

So far we assumed that the leaf $L$ has trivial holonomy, implying
that there is a global set of Casimir functions $C^I$ such that
$L=(C^I)^{-1}(0)$. In other words, the conormal bundle
$(NL)^*=\bigcup_{X\in L}\{\alpha\in T_X^*M: \alpha(v)=0\mbox{ for all
}v\in T_XL\}$ of $L$ is a trivial vector bundle with global basis
$\{\md C^I\}_{I=1\ldots k}$. As seen in Prop.\ \ref{sol2}, for a fixed
leaf $L$ solutions $A$ to the Poisson Sigma Model are then given in
terms of closed 1-forms $\alpha=\alpha_I \, \bard C^I$ on $\Sigma$,
taking values in the pull back of the conormal bundle of $L$.  Here we
denoted the basis in the pull back bundle which corresponds to $\md
C^I$ by $\bard C^I$; it is not to be confused with the pull back of
$\md C^I$, which would be identically zero (and also a section in a
different bundle). The respective contribution to (\ref{solu}) then
can be understood in the following way: Take $\alpha$ as a section in
$T^*\S \otimes \CX^* (NL)^*$, i.e.~$\a \in
\O^1(\S,\CX^* (NL)^*)$. $\CX^* (NL)^*$ is embedded canonically into
$\CX^* T^*M$, we thus may view $\alpha$ also as a particular section
through that bundle. Then we can contract $\alpha$ with $\partial_i$,
viewed as a basis in $\CX^* TM$. Thus the second part of (\ref{solu})
can be written as $\langle \partial_i , \alpha \rangle$. Up to gauge
transformations, only equivalence classes $[\alpha]\in
H^1(\Sigma,\CX^*(NL)^*)$ are representatives. This will be made more
precise in the more general setting to follow.

If the leaf $L$ has non-trivial holonomy, its conormal bundle is
non-trivial and (\ref{solu}) cannot be used as a global
expression. Instead, we have to choose a covering of the leaf such
that in any neighborhood $\CU_i$ of the covering there exist Casimir
functions $C^I_{(i)}$ specifying $L\cap\CU_i=(C^I_{(i)})^{-1}(0)$. If
two neighborhoods have non-empty intersection, there are different
sets of Casimir functions which are related by a transformation
$C^I_{(i)}=f_{(ij)}{}^I_J(C_{(j)})C^J_{(j)}$ as discussed in the
previous subsection. 

\smallskip

\noindent{\it Example:\/} Let $M=[-1,1]^3/\sim$ where the
identifacation $\sim$ is defined by $(1,y,z)\sim (-1,-y,z)$ for all
$y,z\in [-1,1]$, equipped with the Poisson tensor
$\CP=\partial_x\wedge\partial_z$ admitting the compatible
presymplectic form $\wtO=\md x\wedge\md z$. Any section
$z=\mbox{const}$ is a M\"obius strip. The set $L\colon y=0$ is a leaf
in $(M,\CP)$ (while a set $y=\mbox{const}\not=0$ is only half of a
leaf) which can be covered by two neighborhoods ${\cal U}_{1,2}$
admitting the local Casimir function $C=y$. On the full leaf, however,
$y$ is not a global Casimir function since values $y=c$ and $y=-c$
belong to the same leaf for any constant $c\in [-1,1]\backslash
\{0\}$. Correspondingly, we have non-trivial transition functions
$f_1=1$ and $f_2=-1$ in the two intersections of the neighborhoods.

\smallskip

In any neighborhood, (\ref{solu}) is the local expression of a
solution with $(\alpha_I\, \bard C^I)_{(i)}$ representing the solution
as local section of the conormal bundle (plus the cotangent bundle of
$\Sigma$ certainly). The transformation between
different charts is done via (\ref{transC}) such that
$(\alpha_I'\, \bard C^{\prime I})_{(i)}=(\alpha_I\,\bard C^I)_{(j)}$
in ${\cal U}_i\cap{\cal U}_j$. As a global object, therefore, the
local sections $(\alpha_I\,\bard C^I)_{(i)}$ form again a 1-form
$\alpha$ on $\Sigma$ taking values in the pull back of the conormal
bundle of $L$. 

Every local section has to be closed according to the field equations,
and they combine to a global 1-form which is closed in the following
sense: The transition functions of a non-trivial conormal bundle are
given by the {\em constants\/} $f_{(ij)}{}^I_J$, implying that the
conormal bundle is a flat vector bundle\footnote{We are grateful to
A.~Kotov for pointing this out to us.} with a canonical derivative
operator $D$. To be more explicit, one may define an
operator $D$ which annihilates any local basis $(\bard C^I)_{(i)}$;
this derivative is then extended to forms on $\S$ with values in
$\CX^*(NL)^*$ by the graded Leibniz rule: For
$\b \in \Omega^p(\S,\CX^*(NL)^*)$ locally we have 
$\b=(\b_I \bard C^I)_{(i)}$ and then simply $D \b = (\md \b_I \bard
C^I)_{(i)}$.  The locally
$\md$-closed 1-forms $(\alpha_I\, \bard C^I)_{(i)}$ combine to a
globally $D$-closed section $\alpha \in \O^1(\S,\CX^*(NL)^*)$, 
which represents a solution to the field equations. Clearly, $D^2 = 0$
(since the connection is flat, by construction) and there is a natural
cohomology defined on  $\O^p(\S,\CX^*(NL)^*)$, denoted by
$H^p(\S,\CX^*(NL)^*)$. 

To find unique
representatives, we have to consider the symmetry
transformations. As before, the local expressions $(\alpha_I\, \bard
C^I)_{(i)}$ can be changed by adding an exact local 1-form
$(\md\epsilon_I\bard C^I)_{(i)}$. Again, the local 1-forms combine
to a $D$-exact 1-form $D(\epsilon_I\bard C^I)$ taking values in the
pulled back conormal bundle. Thus, solutions are classified by the
first $D$-cohomology $H^1(\Sigma,\CX^*(NL)^*)$.

A similar strategy can be used to deal with leaves which do not admit
a compatible presymplectic form: those leaves can be cut into parts
each admitting a compatible presymplectic form which have to be glued
together by the transformation of the preceding subsection. This will
then imply a transformation (\ref{transO}) for the $\alpha_I$ on
overlapping charts. Returning back to already existing charts then
leads to restrictions on the permitted $\alpha$s. This may lead also
to compactifcations in the solution space, which in previous cases was
always non-compact while it would be compact for, e.g., 2d
$BF$-theories with compact gauge groups. In the present paper,
however, we do not intend to work this out in more detail. 

\subsubsection{Almost topological models}
\label{sec:lsgalmost}

Adding a term \re{nontop}) to the action changes the field equation
for $A$, so we will also obtain different solutions. However, the
changes are not too drastic. To show this we consider the slightly
more general case of an additional term \be
\label{nontopsum}
\int_{\S}C^\s(X(x))\, \varepsilon_\s \ee where a sum of such terms
with possibly different Casimir functions $C^\s$ appears. The
contribution of this addition to the field equations has been
determined in (\ref{eqsnontop}). Using a set of $k$ functionally
independent Casimir functions $C^I$, all the $C^\s$ can be expressed
in terms of these functions (at least locally). Then $C^\s,_i =
C^\s,_I \, C^I,_i$; introducing $\varepsilon_I := \varepsilon_\s
C^\s,_I$ (a 2-form with values in the conormal bundle; we will make
use of this observation below), using \re{eqsnontop}), the key
equation \re{dA}) in the preceding proof is changed to
\begin{equation}
 \md(A_{\underline{i}}+
 \CX^\ast(\partial_{\underline{i}}\lrcorner\wtO))= -\varepsilon_I
 \CX^*(\partial_{\underline{i}} C^I)
\end{equation}
while Eq.\ \re{Aalpha}) still holds true due to
$\partial_{\alpha}C^I=0$. Let us first assume that $\varepsilon_\s =
\md p_\s$  (if some $\epsilon$ are taken to be
a volume form as in Yang--Mills theories, this can be the case only if
$\Sigma$ is non-compact); then also $\varepsilon_I=\md p_I$ is exact
(since $C^\s,_I$ in $p_I = p_\s C^\s,_I$ is, as a function of
Casimirs, constant on $\S$). With the assumption,
any solution $A$ can still be cast into the form \re{solu}), but now
the $\alpha_I$ are not necessarily closed but only $\alpha_I+p_I$ (by
assumption not all $p_I$ can be closed, since otherwise all
$\varepsilon_I=\md p_I$ would vanish). Noting that the symmetries are
unaltered, we obtain

\begin{cor}
  For a given map $\CX$ with image in a symplectic leaf $L$ of trivial
  holonomy which has a neighborhood $U$ permitting a presymplectic
  form $\widetilde \O$ compatible with $\CP$, any solution to the
  field equations \re{eqs1}), \re{eqsnontop}) with $\varepsilon_I=\md
  p_I$ may be written in the form \re{solu}) where $C^I$,
  $I=1,\ldots,k$, are some Casimir functions with $L = (C^I)^{-1}(0)$.
  The 1-forms $\alpha_I+p_I$ are closed on $\S$. For fixed $\widetilde
  \O$ and $C^I$, redefining $\a_I$ by adding an exact 1-form on $\S$
  is a gauge transformation. For a fixed map $\CX$, gauge equivalence
  classes of solutions to the field equations for $A$ correspond to
  the set of $k$ elements $[\a_I+p_I] \in H^1(\S)$.
\end{cor}

If an $\varepsilon_I$ is not exact, we have to proceed more
carefully. For an exact $\varepsilon_I=\md p_I$ we have just seen that
the $\alpha_I$ have to fulfill $\md \alpha_I=-\md
p_I=-\varepsilon_I$. Locally, this will still hold for a non-exact
$\varepsilon_I$ as a consequence of the field equations, but there will be
no 1-forms $\alpha_I$ which can fulfill this equation globally.

For simplicity we first discuss the case $k=1$, i.e.~that there is only
one $\alpha$ which locally fulfills $\md\alpha=\epsilon$. If we choose
a good cover $\{\CU_m\}$ of $\Sigma$ (the neighborhoods $\CU_m$ as
well as their nonvanishing intersections are topologically trivial),
then in each $\CU_m$ we have a 1-form $p_m$ such that $\md
p_m=\epsilon$. Furthermore, due to $\md(p_m-p_n)=\epsilon-\epsilon=0$,
we have functions $\l_{mn}$ with $p_m-p_n=\md \l_{mn}$ on $\CU_m \cap
\CU_n$. Now, $\alpha$ has to fulfill $\md\alpha_m=\epsilon=\md p_m$
which implies $\alpha_m=p_m+\md \l_m$ for some functions $\l_m$ which
can be chosen to be zero by redefinition of $p_m$ or by
using an appropriate local gauge
transformation. In the intersection of two neighborhoods $\CU_m$ and
$\CU_n$,
the local 1-forms of $\alpha$ do not necessarily agree but differ by
an exact form: $\alpha_m-\alpha_n=\md \l_{mn}$. In other words,
$\alpha$ is a connection on the line bundle with curvature $\epsilon$
which is obtained as the pull back of the conormal bundle of the leaf
$L$. As usually, we have $(\alpha-\alpha')_m-(\alpha-\alpha')_n=0$
such that the difference of two such connections $\alpha$ and
$\alpha'$ is a global 1-form. It has to fulfill
$\md(\alpha-\alpha')=0$, i.e., it is closed. Therefore, the space of
all connections of curvature $\epsilon$ can be identified with the
space of closed 1-forms, and the space of gauge equivalence classes
with the first cohomology $H^1(\Sigma)$.

If $k>1$, we obtain the $\alpha_I$ as $k$ connections on $k$ line
bundles the $I$-th one of which has curvature
$\varepsilon_I$. Alternatively, the $\alpha_I$ together can be viewed
as components of an $\fra$-connection on the pull back of the conormal
bundle of the leaf $L$, where $\fra$ is the transversal Lie algebra of
the leaf. In fact, the part $\alpha:=\alpha_IC^I{},_{i}\bard X^i$ of
(\ref{solu}), $\bard X^i$ denoting a basis in $\CX^*T_X^*M$, in a
point $X\in L$ always takes values in the transversal Lie algebra
$\fra_X$ which, as a manifold, can be identified with the conormal
space $(N_XL)^*$.  Furthermore, the transversal Lie algebra of a
regular leaf is always abelian and, therefore, isomorphic to $\dR^k$
with $k={\rm codim}(L,M)$ which coincides with our result of $k$
abelian connections on $k$ line bundles. For a regular leaf the
reformulation via the transversal Lie algebra is thus almost trivial,
but we will see later that it is helpful for a possible generalization
to non-regular leaves.

Finally, gauge transformations of the Poisson Sigma Model have already
been seen to add   exact 1-forms to $\alpha_I$,
which agrees with the notion of gauge transformation for a
connection. Together with the known classification of inequivalent
bundles with connection we obtain

\begin{prop}\label{solnontop2}
  For a given map $\CX$ with image in a symplectic leaf $L$ of trivial
  holonomy which has a neighborhood $U$ permitting a presymplectic
  form $\widetilde \O$ compatible with $\CP$, any solution to the
  field equations \re{eqs1}), \re{eqsnontop}) may be written in the
  form \re{solu}) where $C^I$, $I=1,\ldots,k$, are some Casimir
  functions with $L = (C^I)^{-1}(0)$.

  The $\alpha_I$ form a transversal Lie algebra valued connection on
  the pull back of the conormal bundle of $L$ to $\Sigma$ with
  curvature $\varepsilon_I$. For a fixed map $\CX$, gauge equivalence
  classes of solutions to the field equations for $A$ correspond to
  inequivalent connections of the given curvature on the given line
  bundle on $\Sigma$. All those connections are classified by $k$
  elements of $H^1(\S)$.
\end{prop}

This agrees with the previous results in the case of vanishing or
exact $\varepsilon_I$, in which case the bundles with connection over
$\Sigma$ are trivial. If $L$ has non-trivial holonomy, we can combine
Prop.\ \ref{solnontop2} with the result of the previous subsection. On
each chart $\CU_m$ there is a 1-form $\alpha_m$ with values in the
pull back of the conormal bundle (restricted to the chart). Globally
there is some 2-form $\varepsilon = \varepsilon_I \bard C^I$, taking
values in $\CX^* (NL)^*$, furthermore; on local charts $\{\CU_m\}$, it
has primitives $p_m$, i.e.~$\varepsilon = D p_m$.  Similarly to before
we find $D \a_m = D p_m$, concluding $\a_m = p_m + D \l_m$. With
$\l_{mn} = p_m - p_n$ we then obtain on intersections \be \a_n = \a_m
+ D\left(\l_{mn} +\l_m - \l_n \right) \, , \ee where $\l_m$ reflects
the ambiguity in the definition of the local primitives $p_m$ or
likewise the local gauge freedom.  This defines a kind of connection
on the conormal bundle with ``curvature'' $\varepsilon = \varepsilon_I
\bard C^I$ (on each chart we have $\varepsilon_m \equiv
\varepsilon|_{\CU_m} = D \a_m$). It would be interesting to clarify
the precise mathematical nature of such an object $\{ \a_m \}$.  Since
the $\l_m$ can be gauged to zero and the $\l_{mn}$ are fixed by the
2-forms $\varepsilon_I$ (up to the previously mentioned ambiguity
given by $\l_m$), the \emph{difference} between two such collections
$\{ \a_m \}$ again defines a global 1-form with values in $\CX^*
(NL)^*$. Thus the space of all inequivalent $\alpha$ of the given
``curvature'' $\varepsilon$ is classified by the first cohomology of
conormal bundle valued forms.

Despite the fact that the addition of \re{nontop}) spoils the
topological nature of the model, Corollary \ref{solnontop1} and
Proposition \ref{solnontop2} show that the moduli space of classical
solutions is parameterized by the same topological objects which
classify solutions of the topological models.

\subsubsection{Summary}

Let us summarize the results of this section in

\begin{theo}\label{Summ}
 Let $\Sigma$ be a two-dimensional manifold and $(M,\CP)$ a Poisson
 manifold.

 For stationary points of a topological or almost topological Poisson
 Sigma Model with $\Sigma$ and $(M,\CP)$ the image of the map
 $\CX\colon\Sigma\to M$ is contained in a symplectic leaf $L$ of
 $M$. If $L$ admits a compatible presymplectic form, the space of
 corresponding solutions for $A$ is given by the first cohomology
 class $H^1(\Sigma,\CX^*(NL)^*)$ of forms on $\Sigma$ taking values in the
 pull back via $\CX$ of the conormal bundle of the leaf $L$.  A local
 representation of $A$-solutions is given by (\ref{solu}).
\end{theo}

We already remarked on a possible generalization to cases where a
compatible presymplectic form does not exist globally
(Sec.~\ref{sec:Gen}). Later we will in particular discuss the case of
non-regular leaves.

\section{Examples}
\label{sec:ex}

Applying Theorem \ref{Summ}, we see that in all cases where all leaves
have the same codimension $k$ the solution space $\Mod(\S)$ is of
dimension
\[ 
 \dim\Mod(\S)=k\left(\rank H^1(\S)+1\right)
\] 
which generalizes formula (\ref{dimM}) for the dimension in the
topologically trivial case. However, in the general case the solution
space will not be a linear space because there may be non-trivial
identifications, which depend on the topology of $\S$ and the leaf $L$
and can even lead to a non-Hausdorff topology by gluing the sectors
corresponding to different homotopy classes of maps $\CX\colon\S\to
L$. Specializing Theorem \ref{Summ} to the topologically trivial case
dealt with in Sec.\ \ref{Trivial} shows that we get back the explicit
solutions given there. But Theorem \ref{Summ} is applicable to a class
of Poisson Sigma Models more general by far.  E.g.\ if $M$ is foliated
trivially but by topologically non-trivial leaves provided only their
second cohomology vanishes, all solutions are given by Theorem
\ref{Summ} owing to Lemma \ref{pot}.

We can also compare with the results obtained in Sec.\ \ref{NonAbSol}
for non-abelian $BF$-theories. In this case $M$ is the Lie algebra
$\frg$ of a semisimple Lie group $G$ equipped with the Poisson tensor
$\CP^{ij}=f^{ij}\mbox{}_kX^k$ and the symplectic leaves are identical
to the adjoint orbits in $\frg$ (we identify the Lie algebra of a
semisimple Lie group with its dual by means of the Cartan--Killing
metric).  As compared to Sec.\ \ref{NonAbSol}, we are now solving the
field equations in the opposite direction, i.e.\ we first solve for
$X$. According to Theorem \ref{sol1}, all equivalence classes of
solutions are given by homotopy classes of maps $\CX\colon\S\to L$ for
any leaf $L$. This is identical to the results found in Sec.\
\ref{NonAbSol}. Now, given a solution $X$, solutions for $A$ are given
by Theorem \ref{Summ} in those cases in which a compatible
presymplectic form exists. This can be the case only for non-compact
$G$: for compact $G$ all leaves in $M$ are compact symplectic
manifolds which necessarily have non-trivial second homology, for
otherwise their symplectic form would be exact and so the symplectic
volume would vanish. Now appealing to Corollary \ref{triv} shows that
there is no compatible presymplectic form in any neighborhood of a
given regular leaf because the symplectic volume of any non-trivial
two-cycle is not constant along the direction $\partial_C$ given by
the Casimir function $C(X)=\tr(X^2)$.  Recall that for compact $G$ the
solution space, i.e.\ the space of flat connections, is compact, which
also demonstrates that in this case our methods cannot be applicable
(Theorem \ref{Summ} always implies a non-compact solution space). As
discussed at the end of Sec.~\ref{sec:Gen}, solutions for leaves which
do not admit a global presymplectic form can be found by gluing
solutions obtained with local forms. The gluing procedure will lead to
additional identifications which can compactify the solution space.

If there is a compatible presymplectic form for a non-compact group
$G$ (if, e.g., all leaves have trivial second cohomology, cf.\ Lemma
\ref{pot}), we can apply Theorems \ref{sol1} and \ref{Summ} in order
to find solutions. Solutions for $X$ are given by maps $\CX\colon\S\to
M$ with image contained in an adjoint orbit which coincides with the
observations in Sec.\ \ref{NonAbSol}. However, general results about
the existence of compatible presymplectic forms are available only for
a non-degenerate leaf such that Theorem
\ref{Summ} can directly only lead to solutions with reducible
connections for a semisimple group $G$. In fact, in simple cases one
can show easily that a connection of the form (\ref{solu}) is
reducible: If $\rank G=1$ the Casimir function is $C=\tr(X^2)$,
denoting the Cartan--Killing norm on $\frg$ by $\tr$, and (\ref{solu})
takes the form (using generators $T^i$ of $G$)
\[ A_iT^i=-T^i\CX^*(\partial_i\lrcorner\wtO)+2\alpha X
\] with a closed 1-form $\alpha$ on $\S$. If $\CX$ can be gauged to be
a constant map, the first term vanishes leading to $A=2\alpha X$. This
implies that all holonomies of $A$ are given by $\exp cX$ for some
$c\in\dR$ which shows that $A$ is reducible. We can, therefore, expect
to have access to the generic part of the solution space only if
$\rank\pi_1(\S)$ is small (see, however, possible generalizations
discussed in the next section). Otherwise, the
solution space would be dominated by irreducible connections which
lead to $X$-solutions in the degenerate leaf given by the origin. For
$\rank\pi_1(\S)\leq1$, which physically is most interesting, Theorem
\ref{sol2} determines the generic part of $\Mod(\S)$ because there are
no irreducible connections. In fact, the dimensions of the solution
spaces given in Sec.\ \ref{NonAbSol} and Theorem \ref{Summ} coincide:
in both cases we need $k=\dim\Kern\CP$ parameters to specify a leaf,
which in turn determines the equivalence class of an $X$-solution (up
to certain discrete labels which we need in order to fix the homotopy
class of the map $\CX$), and $k\rank H^1(\S)$ parameters to specify
the $A$-solution. Note that $k=\dim\Kern\CP=\dim G/{\rm Ad}$ so that
the dimensions of the space of reducible flat connections and of the
solution space according to Theorem \ref{Summ} in fact coincide.

Noting that, as already remarked in Sec.\ \ref{NonAbSol}, $X$-solutions
lying in regular leaves correspond to reducible connections whose
holonomies generate a maximal {\em abelian\/} subgroup of $G$, we can
clarify the appearance of $H^1(\S)$ in the Poisson Sigma Model
classification of $A$-solutions as opposed to $\pi_1(\S)$ in the gauge
theory classification: Since all holonomies commute, only the
abelianization of $\pi_1(\S)$ matters, which is just $H^1(\S)$.

Corollary \ref{solnontop1} and Proposition \ref{solnontop2} in
particular provide solutions for Yang--Mills theories when we choose a
quadratic Casimir $C$ and volume form $\epsilon$. Proposition
\ref{solnontop2} only applies if the $X$-solution maps $\Sigma$ into a
non-degenerate leaf, so that we obtain solutions with non-vanishing
electric field $X$ leading to a non-flat connection. The difference
between $BF$- and Yang--Mills theories is automatically accounted for
by the appearance of $\epsilon$ in the conditions for a solution $A$.

\section{Non-regular leaves}
\label{sec:nonregular}

Since we are not aware of general results concerning the existence of
compatible presymplectic forms for non-regular leaves, Theorem
\ref{Summ} does not give us direct access to solutions in this
case. The comparison with non-abelian $BF$-theories shows that in
general we cannot expect non-regular leaves to contribute only a
lower-dimensional set to $\Mod(\S)$; in fact those leaves usually
correspond to solutions forming a dense subset of the moduli
space. Only if the rank of the fundamental group of $\S$ does not
exceed one is the moduli space dominated by solutions corresponding to
non-degenerate leaves. As we will see below, this holds true also for
non-linear Poisson structures.

But the information we obtain is of interest also in cases where
solutions for regular leaves do not correspond to the generic part of
$\Mod(\S)$ in a given model and complements methods which are targeted
to the generic part (e.g.\ the theory of irreducible flat connections
on compact Riemann surfaces used in two-dimensional non-abelian gauge
theories).

The case of gravitational models is special because we have an
additional condition which requires the metric constructed from $A$ to
be non-degenerate. Investigations with other methods
\cite{TK3,TKkinks} suggest that this reduces the contributions from
non-regular leaves such that the methods developed here can have access to
the main part of the moduli space.  In the present paper, however, we
will not dicuss this issue further and instead focus on a possible
generalization of the classification of solutions to non-regular
leaves. 

When discussing the solutions for almost topological Poisson Sigma
Models we already observed that the connection has to be transversal
Lie algebra valued, which in the case of regular leaves is always an
abelian algebra.  To generalize this result we first recall how the
transversal Lie algebra of a point $X\in L$ of a leaf $L$ can be
constructed \cite{SilvaWeinstein}:

\begin{defi}
 The {\em transversal Lie algebra\/}
 of a point $X$ in a leaf $L$ of a
 Poisson manifold $(M,\CP)$ is the conormal space
 $\fra_X:=(NL_X)^*=(T_XM/T_XL)^*\equiv (T_XL)^0=\{\alpha\in
 T^*_XM:\alpha(v)=0 \mbox{ for all }v\in T_XL\}$, identified with the
 annihilator of the tangent space $T_XL$, with the following Lie
 bracket: For two elements $\alpha,\beta\in\fra_X$ we choose functions
 $f$ and $g$ which vanish in a neighborhood of $X$ in the leaf $L$
 such that $\md f_X=\alpha$ and $\md g_X=\beta$. The bracket
\begin{equation}
 [\alpha,\beta]_X:=\md\{f,g\}_X
\end{equation}
 is then well defined and defines the transversal Lie algebra $\fra_X$.
\end{defi}

We will later use another way to identify $\fra_X$ as a submanifold of
the cotangent bundle of $M$:

\begin{lemma} \label{PoissKern}
 As a manifold, the transversal Lie algebra $\fra_X$ is the kernel of
 the Poisson tensor $\CP$ in $X$.
\end{lemma}

\begin{proof}
 For any cotangent vector $\omega\in T^*_XM$ the vector
 $v:=\CP^{\#}(\omega)$ is tangential to $L$ such that
 $\CP(\alpha,\omega)=\alpha(v)=0$ for all $\omega\in T_X^*M$ proving
 that $\fra_X$ is contained in the kernel of $\CP$. Equality of the
 vector spaces then follows from a dimensional argument.
\end{proof}

We are going to discuss the transversal Lie algebra for gauge theories
where $M=\frg^*$ as in Sec.~\ref{NonAb}.

\begin{lemma} \label{TransLieDual}
 If $M=\frg^*$ is the dual of a Lie algebra, then the transversal
 Lie algebra $\fra_X$ of a point $X\in M$ is the isotropy algebra of the
 co-adjoint action of $\frg$ at $X$.
\end{lemma}

\begin{proof}
 As a subspace of the cotangent bundle of $\frg^*$, $\fra_X$ is
 naturally identified with a subspace of
 $\frg^{**}\equiv\frg$. Furthermore, it follows from the definition
 that $\fra_X$ is also a subalgebra of $\frg$: if $\alpha=\md
 f_X=f,_{i}\md X^i$ and $\beta=\md g_X=g,_{i}\md X^i$ are in the
 kernel of $\CP_X^{\sharp}$, we have
\[
 \md \{f,g\}=\md (f^{ij}{}_{k}X^kf,_{i}g,_{j})=
 f^{ij}{}_kf,_{i}g,_{j}\md X^k
\]
 which implies
\[
 [\alpha,\beta]_X=\md\{f,g\}=f^{ij}{}_k \alpha_i\beta_j\md X^k\equiv
 [\alpha,\beta]_{\frg}
\]
 where the last bracket denotes the usual bracket in the Lie algebra
 $\frg$ and $\md X^k$ are identified with the generators of
 $\frg^{**}\equiv\frg$.

 The condition for $\fra_X$ of Lemma \ref{PoissKern} now reads
\[
 \CP_X(\alpha,\omega)=X([\alpha,\omega]_{\frg})=({\rm
 coad}_{\alpha}X)(\omega)=0
\]
 for all $\omega\in\frg$ which concludes the proof.
\end{proof}

If $\frg$ is semisimple, we can identify $M=\frg^*$ with $\frg$ and
the point $X\in M$ with an element of $\frg$. The transversal Lie
algebra $\fra_X$ then is the subalgebra of $\frg$ fulfilling
$[X,\alpha]=0$ for all $\alpha\in\fra_X$.

We are now ready to exploit this information in the context of
solutions to gauge theories. We already know that solutions for the
field $X$ are given by arbitrary maps of $\S$ into a leaf $L$ of
$M$. Locally, the map ${\cal X}$ can be deformed by gauge
transformations such that its image is a single point $X\in L$;
therefore, the field equation $\md X+[A,X]=0$ implies that all
solutions for $A$ have to commute with $X\in\frg$ and thus, according
to Lemma \ref{TransLieDual}, have values in the transversal Lie
algebra of $X$. Furthermore, the local component 1-forms $A_i$ such
that $A=A_i\md X^i$ of all those connections can be written as
$\alpha_IC^I{},_{i}$ with $k$ 1-forms $\alpha_I$ where $k$ is the
codimension of the leaf and $C^I$ are $k$ Casimir functions specifying
the leaf. If the image of $X$ is not just a single point, we need an
additional contribution $a$
for the connection such that $[a,X]=-\md X$. Then,
$A:=a+\alpha_IC^I{},_{i}T^i$ would provide a solution to $\md
X+[A,X]=0$. If we can find such a form $a$, any solution to the
first field equation can be written as $a$ plus a transversal Lie
algebra valued connection. This demonstrates that the role of the
transversal Lie algebra is unchanged if we have a non-regular
leaf. Now it is easy to see that $a=-\partial_i\lrcorner\wtO T^i$
with $\wtO$ compatible with $\CP$ is appropriate because
\[
 [\partial_i\lrcorner \wtO T^i,X]=f^{ij}{}_k\wtO_{il}\md X^l X_j
 T^k= \md X^k T_k
\]
provided that $f^{ij}{}_kX_j\wtO_{i\alpha}= -\CP^{ik}\wtO_{i\alpha}=
\delta^k_{\alpha}$ (where $\alpha$ can be regarded as a tangential
index since it is contracted with $\md X^l$). Note that for this
equation $\wtO$ only needs to give the leaf symplectic structure when
restricted to the leaf itself which is weaker than the condition for a
compatible presymplectic form. However, one also has to assure that
$a$ has the correct curvature in order for $A$ to solve the second
field equation. For regular leaves this requires $a$ to be
constructed with a compatible presymplectic form as we have seen.

The last calculations suggest that the classification of solutions to
Poisson Sigma Models as found in this paper generalizes to arbitrary
leaves where the transversal Lie algebra plays the role of the
connection 1-forms $\alpha_I$. (This is also suggested by a
reinterpretation of the solutions as Lie algebroid morphisms
\cite{bks}.) Only the explicit form
(\ref{solu}) of a solution cannot be used if there is no substitute
for the compatible presymplectic form $\wtO$. One example where one
can easily find an alternative form for a non-regular leaf is the
origin as a degenerate leaf in a semisimple Lie algebra: Here we can
choose $a=0$ since all maps $X$ into this leaf have only one image
point, which would correspond to a form $\wtO$ which is not compatible
with $\CP$ in a neighborhood of the leaf. In this case, the
transversal Lie algebra agrees with the Lie algebra itself such that
all solutions for $A$ are given by Lie algebra valued connections with
the correct curvature (zero for $BF$ theories or given by the volume
form for Yang--Mills theories). Thus, the methods of the present paper
give us the well-known results also for a degenerate leaf, in which
case we obtain irreducible connections and a vanishing $X$.

In fact, this conclusion does not only apply to $BF$ and Yang--Mills,
which have a linear Poisson tensor, but to a general Poisson Sigma
Model as well provided that the image of $\CX$ is contractible in the
leaf $L$. In this case, one can choose the gauge in which the image of
$\CX$ is a single point where the previous remarks can be used. The
$A$-components tangential to the leaf in the given point must be zero
in this gauge owing to the first field equation, while the remaining
components are subject to the second field equation with structure
{\em constants\/} $P^{IJ},_K$ of the transversal Lie algebra. Thus,
up to gauge transformations,
$A$ has to be a flat transversal Lie algebra valued connection
whenever the $\CX\colon \S \to L$ in (\ref{sol1}) is of trivial
homotopy---so that $X$ can be gauged to be constant. The remaining
gauge freedom then gives the usual gauge transformations of a connection.

In this special case, the field equations of a general Poisson Sigma
Model can be reduced to those of $BF$-theory, and also the formulas we
obtained in Sec.~\ref{NonAbSol} for the dimensions of subspaces
$\Mod(\S)$ corresponding to different classes of leaves (regular or
non-regular) can be used. Regular leaves $L_{\rm reg}$ always
contribute solutions which form a subspace of dimension
$(\rank\pi_1(\S)+1){\rm codim}(L_{\rm reg},M)$ while a degenerate leaf
$L_{\rm deg}$ yields $(\rank\pi_1(\S)-2){\rm codim}(L_{\rm deg},M)$.
Since ${\rm codim}(L_{\rm deg},M)>{\rm codim}(L_{\rm reg},M)$ for a
Poisson tensor of non-constant rank, the contribution of a degenerate leaf
will always dominate the solution space provided that the rank of the
fundamental group of $\S$ is large enough. Similarly, one can see that
the dimension of the solution space for any leaf $L$ (not necessarily
regular or degenerate) is approximately given by $\rank\pi_1(\S){\rm
codim}(L,M)$ for large rank of the fundamental group (this is the
contribution of connections taking values in the transversal Lie
algebra of dimension ${\rm codim}(L,M)$, while the contribution of
$X$-solutions is not proportional to the rank of the fundamental group
and thus sub-dominant). Therefore, regular leaves will not give a
dense subset of the solution space for large fundamental group of
$\S$, even when there are no degenerate leaves. In other words, the
leaves of the lowest dimension dominate the solution
space. 

\section*{Acknowledgements}

We thank A.\ Cattaneo, A.\ Kotov, and J.\ Stasheff for
discussions. T.S.\ is grateful to the Erwin Schr\"odinger Institute in
Vienna for hospitality in the period when this work was begun, and
M.~B.\ to A.\ Wipf and the TPI in Jena for hospitality. The work of
M.~B.\ was supported in part by NSF grant PHY00-90091 and the Eberly
research funds of Penn State.


\begin{thebibliography}{10}

\bibitem{PSM1}
P.\ Schaller and T.\ Strobl,
\newblock Poisson structure induced (topological) field theories,
\newblock {\em Mod. Phys. Lett.}, A9:3129--3136, 1994.

\bibitem{Ikeda}
N.\ Ikeda,
\newblock Two-dimensional gravity and nonlinear gauge theory,
\newblock {\em Ann. Phys.}, 235:435--464, 1994.

\bibitem{CFHam}
A.~S.\ Cattaneo and G.\ Felder,
\newblock {P}oisson sigma models and symplectic groupoids,
\newblock math.SG/0003023.

\bibitem{Ctirad}
Ct.\ Klimcik and T.\ Strobl,
\newblock {WZW}-{P}oisson manifolds,
\newblock {\em J. Geom. Phys.}, 43:341--344, 2002.

\bibitem{G/G}
A.~Yu.\ Alekseev, P.\ Schaller, and T.\ Strobl,
\newblock The topological {$G/G$} {WZW} model in the generalized momentum
  representation,
\newblock {\em Phys. Rev.}, D52:7146--7160, 1995.

\bibitem{PSM3}
P.\ Schaller and T.\ Strobl,
\newblock Introduction to {Poisson-$\sigma$} models,
\newblock In H.~Grosse and L.~Pittner, editors, {\em Low-Dimensional Models in
  Statistical Physics and Quantum Field Theory}, volume 469 of {\em Lecture
  Notes in Physics}, page 321 (Springer, Berlin, 1996).

\bibitem{TK1}
T.\ Kl{\"o}sch and T.\ Strobl,
\newblock Classical and quantum gravity in (1+1)-dimensions. {P}art 1: {A}
  unifying approach,
\newblock {\em Class. Quant. Grav.}, 13:965--984, 1996.
\newblock Erratum ibid. 14 (1997) 825.

\bibitem{Habil1}
T.~Strobl,
\newblock {\em Gravity in {T}wo {S}pacetime {D}imensions},
\newblock Habilitationsschrift, Rheinisch-Westf\"alische Technische Hochschule
  Aachen, 1999.

\bibitem{PSM2}
P.\ Schaller and T.\ Strobl,
\newblock {P}oisson sigma models: {A} generalization of 2d gravity
  {Y}ang-{M}ills systems, hep-th/9411163.

\bibitem{StMarg}
T.~Strobl,
\newblock 2d quantum dilaton gravity as/versus a finite dimensional quantum
  mechanical systems,
\newblock {\em Nucl. Phys. Proc. Suppl.}, 57:330--333, 1997.

\bibitem{Woodhouse}
N.~M.~J.\ Woodhouse,
\newblock Geometric quantization, Oxford mathematical monographs,
\newblock New York, Clarendon, 1992.

\bibitem{BayenI}
F.~Bayen, M.~Flato, C.~Fronsdal, A.~Lichnerowicz, and D.~Sternheimer,
\newblock Deformation theory and quantization. 1. deformations of symplectic
  structures,
\newblock {\em Ann. Phys.}, 111:61, 1978.

\bibitem{BayenII}
F.~Bayen, M.~Flato, C.~Fronsdal, A.~Lichnerowicz, and D.~Sternheimer,
\newblock Deformation theory and quantization. 2. physical applications,
\newblock {\em Ann. Phys.}, 111:111, 1978.

\bibitem{Kontsevich}
M.\ Kontsevich,
\newblock Deformation quantization of poisson manifolds, I,
\newblock q-alg/9709040.

\bibitem{CF1}
A.~S.\ Cattaneo and G.\ Felder,
\newblock A path integral approach to the {K}ontsevich quantization formula,
\newblock {\em Commun. Math. Phys.}, 212:591, 2000.

\bibitem{Volker}
V.\ Schomerus,
\newblock D-branes and deformation quantization,
\newblock {\em JHEP}, 06:030, 1999.

\bibitem{SWmap}
N.\ Seiberg and E.\ Witten,
\newblock String theory and noncommutative geometry,
\newblock {\em JHEP}, 09:032, 1999.

\bibitem{Moyal}
J.~E.\ Moyal,
\newblock Quantum mechanics as a statistical theory,
\newblock {\em Proc. Cambridge Phil. Soc.}, 45:99--124, 1949.

\bibitem{brackets}
M.\ Bojowald and T.\ Strobl,
\newblock Poisson geometry in constrained systems,
\newblock hep-th/0112074.

\bibitem{TK2}
T.\ Kl{\"o}sch and T.\ Strobl,
\newblock Classical and quantum gravity in 1+1 dimensions. {P}art 2: {T}he
  universal coverings,
\newblock {\em Class. Quant. Grav.}, 13:2395--2422, 1996.

\bibitem{TK3}
T.~Kl{\"o}sch and T.~Strobl,
\newblock Classical and quantum gravity in (1+1)-dimensions. {P}art 3:
  {S}olutions of arbitrary topology,
\newblock {\em Class. Quant. Grav.}, 14:1689--1723, 1997.

\bibitem{TKkinks}
T.\ Kl{\"o}sch and T.\ Strobl,
\newblock A global view of kinks in 1+1 gravity,
\newblock {\em Phys. Rev.}, D57:1034--1044, 1998.

\bibitem{Park}
J.-S.\ Park,
\newblock Topological open p-branes,
\newblock hep-th/0012141.

\bibitem{3Poisson}
P.\ Severa and A.\ Weinstein,
\newblock Poisson geometry with a 3-form background,
\newblock {\em Prog. Theor. Phys. Suppl.}, 144:145--154, 2001.

\bibitem{HT}
M.~Henneaux and C.~Teitelboim,
\newblock Quantization of gauge systems,
\newblock Princeton University Press, 1992.

\bibitem{PLB}
P.\ Schaller and T.\ Strobl,
\newblock Diffeomorphisms versus nonabelian gauge transformations: An example
  of (1+1)-dimensional gravity,
\newblock {\em Phys. Lett.}, B337:266--270, 1994.

\bibitem{Stasheff}
R.\ Fulp, T.\ Lada, and J.\ Stasheff,
\newblock Noether's variational theorem II and the BV formalism,
\newblock math.QA/0204079.

\bibitem{bks}
M.\ Bojowald, A.\ Kotov, and T.\ Strobl,
\newblock in preparation.

\bibitem{prep}
T.\ Strobl,
\newblock Gravity from Lie algebroid morphisms,
\newblock in preparation.

\bibitem{WeinsteinDarboux}
A.\ Weinstein,
\newblock The local structure of {P}oisson manifolds,
\newblock {\em J. Diff. Geom.}, 18:523--557, 1983.

\bibitem{Goldman}
W.~M.\ Goldman,
\newblock The symplectic nature of fundamental groups of surfaces,
\newblock {\em Adv.\ Math.}, 54:200--225, 1984.

\bibitem{VaismanBook}
I.\ Vaisman,
\newblock {\em Lectures on the Geometry of Poisson Manifolds},
\newblock Birkh\"auser, Basel, 1994.

\bibitem{CFm1}
M.\ Crainic and R.~L. Fernandes,
\newblock {I}ntegrability of {L}ie brackets,
\newblock {\em Annals of Mathematics}, 157:575, 2003.

\bibitem{CFm2}
M.\ Crainic and R.~L. Fernandes,
\newblock {I}ntegrability of {P}oisson brackets, math.DG/0210152.

\bibitem{SilvaWeinstein}
A.~Cannas da~Silva and A.\ Weinstein,
\newblock {\em {G}eometric {M}odels for {N}oncommutative {A}lgebras}, volume~10
  of {\em Berkeley Mathematics Lecture Notes},
\newblock American Mathematical Society, Providence, RI, 1999,
\newblock available at \verb|http://www.math.berkeley.edu/~alanw/|.

\end{thebibliography}
\end{document}